\documentclass[a4paper,11pt]{article}
\pdfoutput=1 

\usepackage{jheppub} 

\usepackage[T1]{fontenc} 
\usepackage{lineno}
\usepackage{physics}
\usepackage{orcidlink}

\newcommand{\pipi}{\pi^+\pi^-}
\newcommand{\KK}{K^+ K^-}
\newcommand{\LL}{\ell^+ \ell^-}
\newcommand{\EE}{e^+e^-}
\newcommand{\pipipi}{\pi^+ \pi^- \pi^0}
\newcommand{\MM}{\mu^+\mu^-}

\newcommand{\jpsi}{J/\psi}
\newcommand{\etac}{\eta_{c}}

\newcommand{\ppbar}{p\bar{p}}

\newcommand{\Xcc}{Y_{cc}}

\newcommand{\gevcc}{{\rm GeV}/c^2}
\newcommand{\mevcc}{{\rm MeV}/c^2}
\newcommand{\gisr}{\gamma_{\rm ISR}}

\title{\boldmath Search for the double-charmonium state with $\etac\jpsi$ at Belle}

\collaboration{The Belle Collaboration}
  \author{J.~H.~Yin\,\orcidlink{0000-0002-1479-9349},} 
 \author{Y.~B.~Li\,\orcidlink{0000-0002-9909-2851},} 
  \author{E.~Won\,\orcidlink{0000-0002-4245-7442},} 
  \author{I.~Adachi\,\orcidlink{0000-0003-2287-0173},} 
  \author{H.~Aihara\,\orcidlink{0000-0002-1907-5964},} 
  \author{S.~Al~Said\,\orcidlink{0000-0002-4895-3869},} 
  \author{D.~M.~Asner\,\orcidlink{0000-0002-1586-5790},} 
  \author{T.~Aushev\,\orcidlink{0000-0002-6347-7055},} 
  \author{R.~Ayad\,\orcidlink{0000-0003-3466-9290},} 
  \author{V.~Babu\,\orcidlink{0000-0003-0419-6912},} 
% ...

\emailAdd{yinjh2012@korea.ac.kr}

\abstract{We measure the cross section of $e^+e^-\rightarrow\eta_c J/\psi$ at the $\Upsilon(nS) (n=1$---$5)$ on-resonance and 10.52 GeV off-resonance energy points using the full data sample collected by the Belle detector with an integrated luminosity of $955~\rm fb^{-1}$. 
We also search for double charmonium production in $e^+e^-\rightarrow\eta_c J/\psi$ via initial state radiation near the $\eta_c J/\psi$ threshold.
No evident signal of the double charmonium state is found, but evidence for the $e^+e^-\rightarrow\eta_c J/\psi$ process is found with a statistical significance greater than $3.3\sigma$ 
near the $\etac\jpsi$ threshold.
The average cross section near the threshold is measured and upper limits of cross sections are set for other regions.

}

\begin{document} 
\maketitle
\flushbottom

\section{Introduction}
Early in this century, a number of exotic states, the so called "$XYZ$" particles, have been discovered~\cite{reviewXYZ} via their decays into two heavy-flavor mesons and/or a quarkonium and one or two light hadrons. 
Vector states with $J^{PC}=1^{--}$, such as the
$\psi(4260)$~\cite{Aubert:2005rm}, $\psi(4360)$~\cite{belle_y4660,babar_y4360}, and
$\psi(4660)$~\cite{belle_y4660}, are alternatively called $Y$ states.
The $\psi(4260)$ state is observed for the first time by the BABAR experiment with a mass of
$(4259\pm8^{+2}_{-6}) {\rm~MeV}/c^2$ using the initial state
radiation (ISR) events $\EE\to\gamma_{\rm ISR} \pi^{+}\pi^{-}
J/\psi$~\cite{Aubert:2005rm}. 
The observation was later confirmed
by CLEO~\cite{cleoY4260} and Belle~\cite{Yuan:2007sj}.
In Ref.~\cite{Chiu:2005ey}, the authors calculated the mass of $\psi(4230)$ as $4238 \pm 31 {\rm MeV}/c^2$ using lattice quantum chromodynamics (QCD) by treating this state as a molecule. 
Moreover, the authors predict two additional exotic states with quark compositions of $cs\bar{c}\bar{s}$ and $cc\bar{c}\bar{c}$ with masses of $(4450\pm100)$ MeV/$c^{2}$ and $(6400\pm50)$ MeV/$c^{2}$, respectively.

In 2017, a dedicated analysis performed by the BESIII experiment
revealed that the so-called $\psi(4260)$ state is not simply one
resonance but two~\cite{Ablikim:2016qzw}. The first, called the $\psi(4230)$ state, 
has a lower mass and a much narrower width, while the second at around $4.32 {\rm~GeV}/c^2$ is observed for the first time
with a significance greater than $7.6\sigma$. The lower-mass
resonance was also observed in $\EE\to\pipi h_c$~\cite{Ablikim:2013wzq}~\cite{czy}, 
$\EE\to\omega\chi_{c0}$~\cite{Ablikim:2019apl} and  $\pi \bar{D}
D^{*}+c.c.$ events~\cite{Ablikim:2018vxx}.
The $\psi(4230)$ state is also observed to have a relative large decay rate to lower charmonium states via $\eta$ transition; \textit{viz} $\psi(4230)\to\eta\jpsi$~\cite{Y4220etaJpsi}, $\psi(4230)\to\eta'\jpsi$~\cite{{Y4220etapJpsi}}.

Recently, BESIII reported the cross section measurements of $\EE\to\KK\jpsi$~\cite{Ablikim:2018epj}\cite{BESIII:2022joj}.
A structure is observed with $M=4487.7\pm13.3\pm24.1~\mevcc$, which is very close to the above prediction of $4450~\mevcc$ for $cs\bar{c}\bar{s}$.
Belle also reported a structure with a mass around $4620~\mevcc$ in the cross section measurements of $\EE\to D_s^+ D_{s1}(2536)^-+c.c.$~\cite{Belle:DsDs1} and $D_s^+ D^{*}_{s2}(2573)^-+c.c.$~\cite{Belle:DsDs2}.
Additionally, LHCb reported the observation of a possible $cs\bar{c}\bar{s}$ state in $B^+\to K^+ \phi \jpsi$ decays~\cite{LHCb:phiJpsi1}\cite{LHCb:phiJpsi2}.
Furthermore, LHCb reported pronounced structures in the invariant mass spectrum of $\jpsi$ pairs~\cite{LHCb_X6900}. 
An enhancement in the near-double-$\jpsi$ threshold region from $6.2$ to $6.8~\rm GeV$ is seen, followed by another narrow peak around $6.9~\rm GeV$, dubbed $X(6900)$. 
The interaction between the two charmonia may not be strong enough to form a tight bound state, many theoretical studies adopt the compact tetraquark picture~\cite{review_chenhx}.
The compact diquark anti-diquark structure $[QQ][\bar Q \bar Q]$ is the most popular one, but the mass predictions are quite model dependent~\cite{cccc_1,cccc_2,cccc_3,cccc_4,cccc_5,cccc_6,cccc_7,cccc_8,cccc_9,cccc_10,cccc_11,cccc_12,cccc_13}.
More experimental and theoretical investigations are crucial to understand them.





The lowest mass combination of charmonia to which a vector $cc\bar{c}\bar{c}$ could decay is $\eta_c J/\psi$, 
and this process may have a relative large branching fraction.
We present the results of a search for such a vector $cc\bar{c}\bar{c}$ state, hereinafter designated $\Xcc$, with the Belle detector~\cite{belle} at the KEKB asymmetric-energy $\EE$ collider~\cite{kekb}.
The integrated luminosity is $980~\rm fb^{-1}$, about 70\% of which were collected at the $\Upsilon(4S)$ resonance; the rest were taken at other $\Upsilon(1,2,3,5S)$ states or center-of-mass (c.m.) energies just below the $\Upsilon(4S)$ or the $\Upsilon(nS)$ peaks by tens of MeV, as well as various c.m. energies between $10.63~\rm GeV$ and $11.02~\rm GeV$.
Initial state radiation (ISR) allows us to search for the double-charmonium state in the near-threshold region.
We first measure the cross section of $\EE\to\etac\jpsi$ on the on-resonance $\Upsilon(nS)$ energy point, 
providing validation for our method as well as a solid check for the next-to-next-to-leading-order calculation in the nonrelativistic QCD approach~\cite{NNLO-cal}.
We then search for the possible $\etac\jpsi$ and $\Xcc$ signals in the near-threshold region.
We extrapolate the measured cross section from the on-resonance points to the near-threshold region to check whether the possible $\etac\jpsi$ signals here are from continuum production.

\section{Belle detector and data samples}
The Belle detector~\cite{belle} is a large-solid-angle magnetic
spectrometer that consists of a silicon vertex detector
(SVD), a 50-layer central drift chamber (CDC), an array
of aerogel threshold Cherenkov counters (ACC), a barrel-like arrangement of time-of-flight scintillation counters (TOF), and an
electromagnetic calorimeter (ECL) consisting of CsI(Tl)
crystals. All these detector components are located inside a superconducting solenoid coil that provides a 1.5 T
magnetic field. An iron flux-return located outside of the
coil is instrumented with resistive plate chambers to detect $K^0_L$ mesons and to identify muons.

The signal  Monte Carlo (MC) samples of the ISR processes and the decays of $\eta_c$ and $J/\psi$ are simulated with the {\sc phokhara}~\cite{Rodrigo:2001kf} and {\sc evtgen}~\cite{evtgen} event generators, respectively, with the decay branching fractions and resonance parameters of $\etac$ and $\jpsi$ taken from Ref.~\cite{PDG}.
These events are processed by a detector simulation based on {\sc GEANT3}~\cite{geant3}.
The generic MC samples, corresponding to six times the integrated luminosity of the data, of $\EE\to\Upsilon(nS)$ events with subsequent $\Upsilon(nS)$ decays and $e^+e^- \to q \bar{q}$ ($q=u,~d,~s,~c$) events are used to check the backgrounds.
A tool named {\sc topoana}~\cite{topoana} is used to visualize the MC event types after event selection.
A series of signal MC samples is generated with different $m(\Xcc)$ assumptions to estimate the signal resolutions as well the c.m. energy-dependent efficiency.

\section{Event selection}

Two distinct reconstruction methods are implemented in this analysis.
One is exclusive reconstruction of $\eta_c\jpsi$, and the other is inclusive reconstruction using $\jpsi$ or $\jpsi\gamma_{\rm ISR}$ for $\Upsilon(nS)$ on/off resonance or near-threshold data.

Charged particle tracks are required to have impact parameters perpendicular to and
along the beam direction with respect to the interaction point
of less than 1.0 and 4.0~cm, respectively.
The transverse momentum of each track is required to be greater than 100 MeV/$c$.
For particle identification (PID), except for tracks from $K^0_S$, information from different detector subsystems is combined to form the likelihood $\mathcal{L}_{i}$ for species $i$, where $i= e,~\mu,~\pi$,~$K$, or $p$~\cite{PID}.
Particles with $\mathcal{L}_{K}/(\mathcal{L}_K+\mathcal{L}_{\pi})>0.6$ are regarded as kaons, while those with a ratio below $0.4$ are identified as pions. 
A track with $\mathcal{L}_{p}/(\mathcal{L}_{p}+\mathcal{L}_\pi) > 0.6$ and $\mathcal{L}_{p}/(\mathcal{L}_{p}+\mathcal{L}_K) > 0.6$ is identified as a proton.
Distinct likelihoods are used for muon~\cite{muID} and electron identification~\cite{eID}: 
we require $\mathcal{L}_{\mu}/(\mathcal{L}_{\mu}+\mathcal{L}_{K}+\mathcal{L}_{\pi})>0.1$ for muon candidates and $\mathcal{L}_e/(\mathcal{L}_e+\mathcal{L}_{{\rm non}-e}) > 0.01$ for electron candidates.
These PID requirements provide a relatively high selection efficiency and a low misidentification rate, i.e., about 80\% and 7\%, respectively, for pions.

The photon with the largest energy in the c.m. frame in an event is taken as the ISR photon; its energy is required to be greater than 1 GeV.
$K^0_S$ candidates are reconstructed by combining two tracks of opposite charge reconstructed using the pion hypothesis and consistent with originating from a displaced vertex.
Combinatorial background is suppressed using a neural network~\cite{NNKs}.
Neutral pion candidates are reconstructed from pairs of photons, each photon having deposited energy of at least 50 MeV in the barrel region of the ECL (polar angle within interval $[32.2^{\circ},~128.7^{\circ}]$), or at least 100 MeV in the end-caps (polar angle within interval $[12.4^{\circ},~31.4^{\circ}] \cup [130.7^{\circ},~155.1^{\circ}]$).
The invariant mass of the $\pi^0$ candidate is required to be within the interval $[0.115,~0.155]~\gevcc$, which encompasses a  $3\sigma$ mass window around the nominal mass.
A mass-constrained fit is performed to each surviving $\pi^0$ candidate to improve its momentum resolution.

Lepton pairs, $\EE$ or $\MM$, are used to reconstruct $\jpsi$.
To reduce the effects of bremsstrahlung and final-state radiation, all photons within a 50 mrad cone of the initial electron or positron direction are included in the calculation of the candidate's four-momentum.
The mass window of $\jpsi$ is optimized to be $M(\EE)\in[3.000,~3.120]~\rm GeV/c^2$ and $M(\MM)\in[3.075,~3.125]~\rm GeV/c^2$ using the figure of merit ${\rm FOM} = N_{\rm S}/\sqrt{N_{\rm B}+0.5}$, where $N_{\rm S}$ is the number of events in signal MC (depending on different mass window requirements, and $N_{\rm B}$ is the number of events in generic MC).
In the exclusive reconstruction, six hadronic channels are used to reconstruct $\etac$: $\ppbar$, $\ppbar\pi^0$, $K^0_S K^{\pm}\pi^{\mp}$, $\KK \pi^0$, $\KK\KK$, $2(\pipipi)$.
The number of extra charged tracks found in the detector must be less than three.
The optimized mass window is $M(\etac)\in[2.78,~3.08]~\rm GeV/c^2$.
The candidate with the smallest value of the mass recoiling against $\eta_c J/\psi$, 
$M^2_{\rm recoil}\equiv | p_{\EE} - p(\etac) - p(\jpsi) |^2/c^2$
is chosen as the best candidate;
here, $p$ is the four-momentum of the specified particle in lab frame.
In the inclusive reconstruction, the number of charged tracks is required to be greater than four to suppress the QED backgrounds.

\section{Data analysis}
\subsection{Double charmonium production at $\Upsilon(nS)$ on-resonance and $\Upsilon(4S)$ off-resonance energies}

\begin{figure*}[htbp]
\begin{center}
    \includegraphics[width=0.30\textwidth]{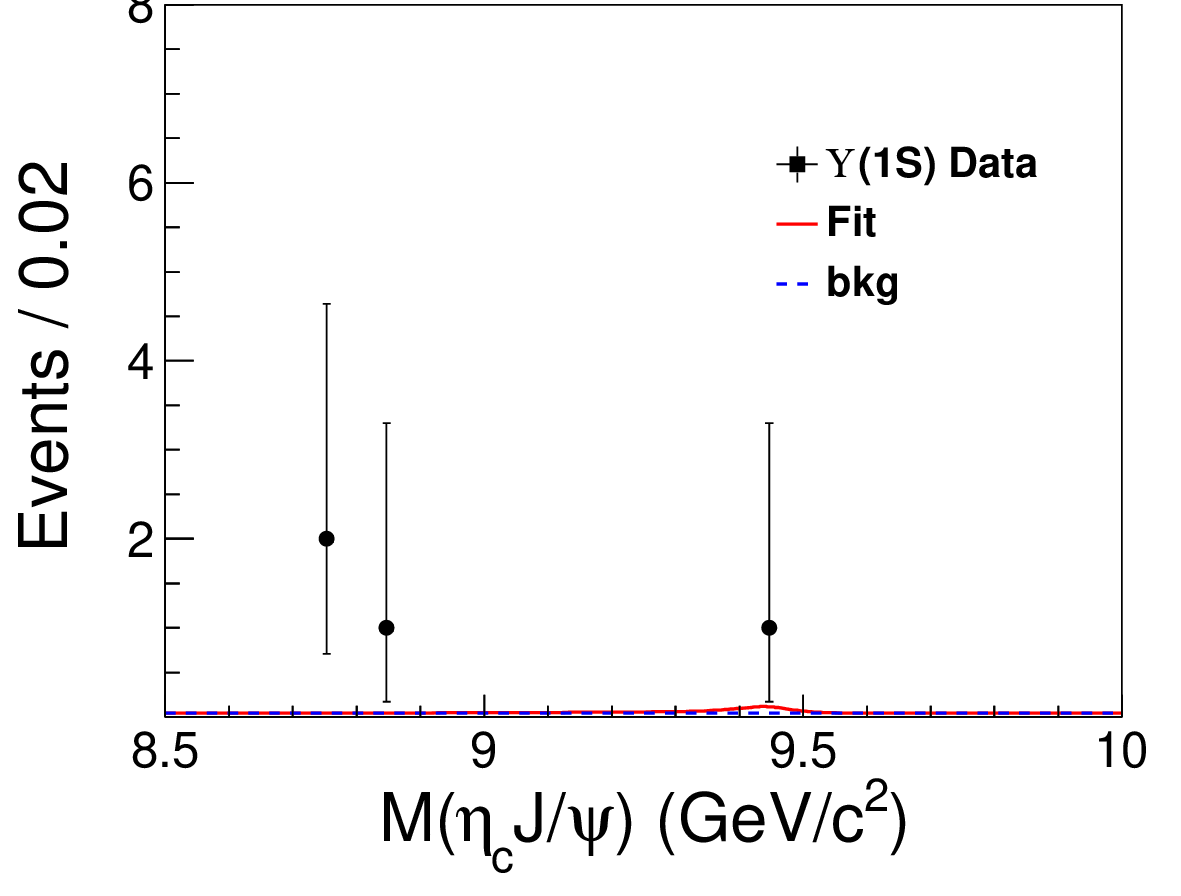} 
    \includegraphics[width=0.30\textwidth]{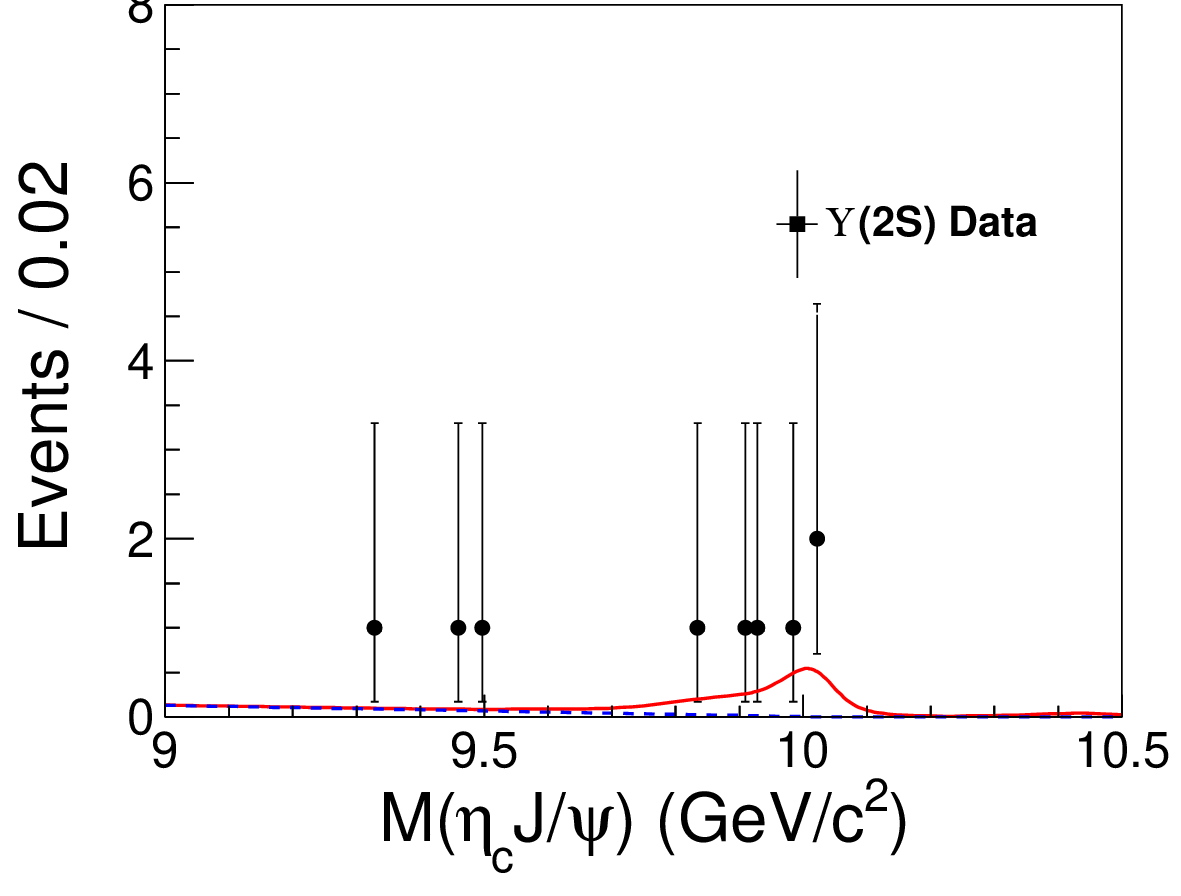} 
    \includegraphics[width=0.30\textwidth]{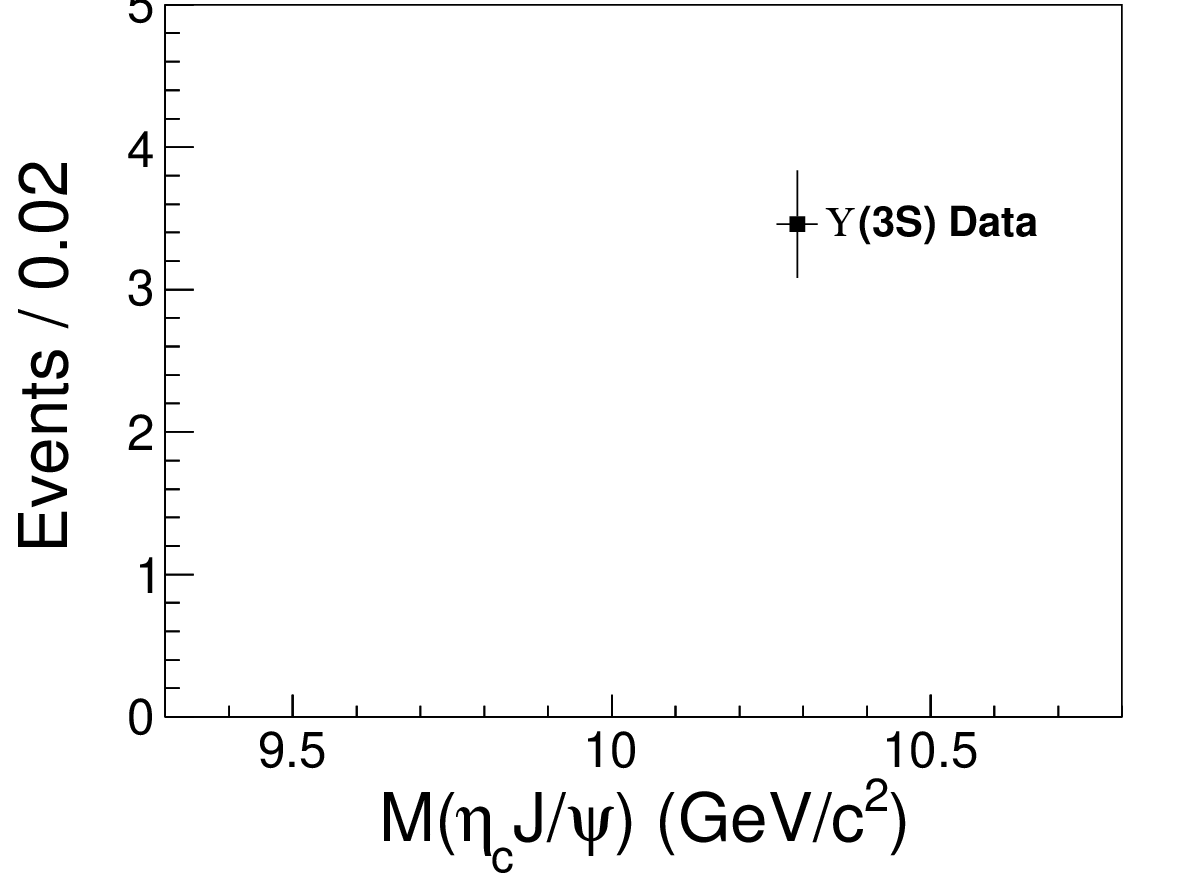} 
    \includegraphics[width=0.30\textwidth]{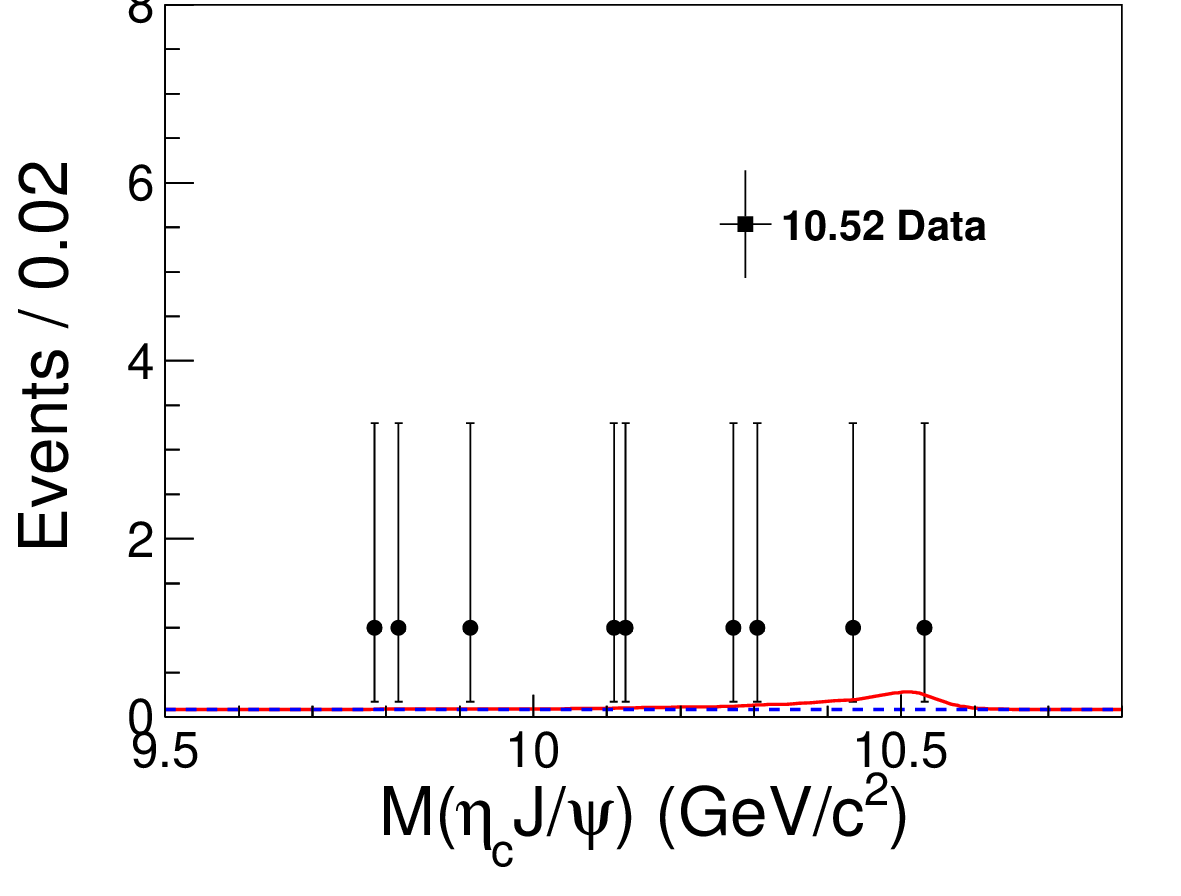} 
    \includegraphics[width=0.30\textwidth]{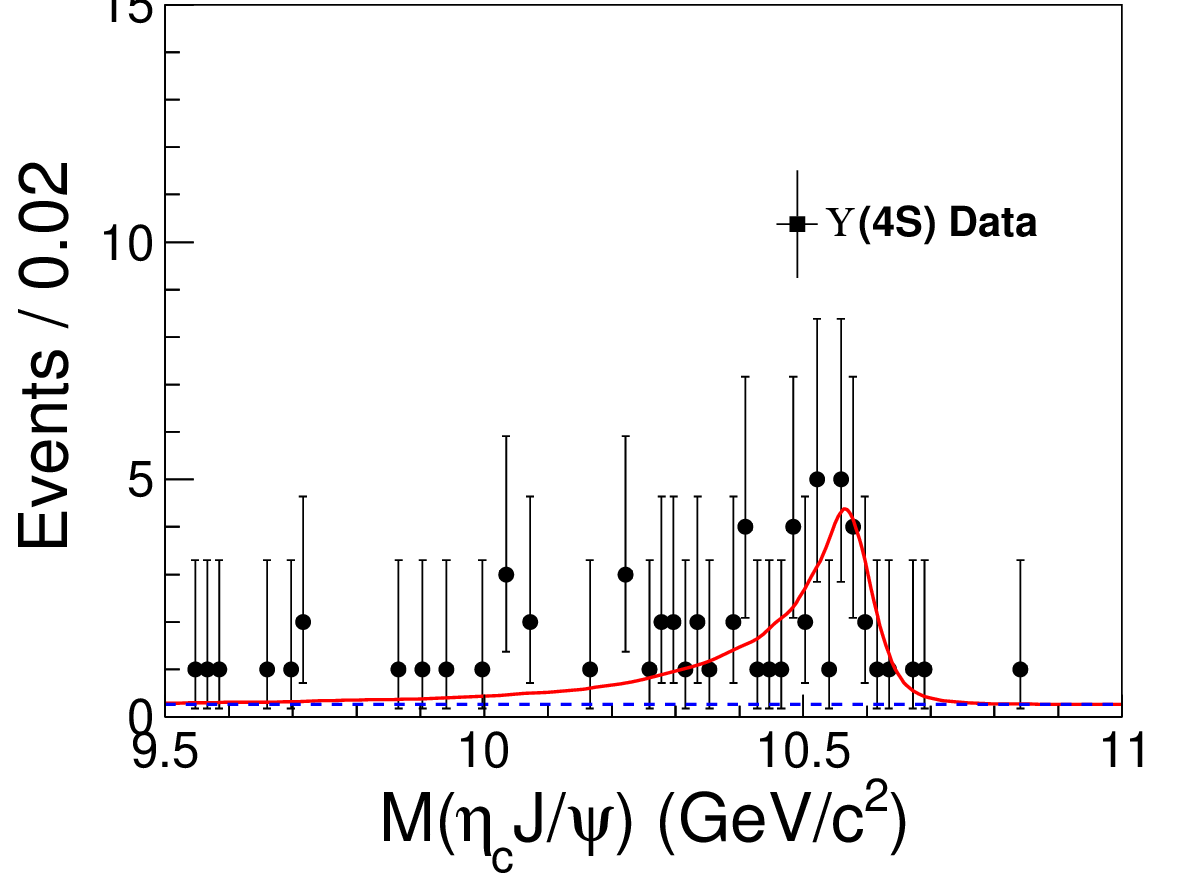} 
    \includegraphics[width=0.30\textwidth]{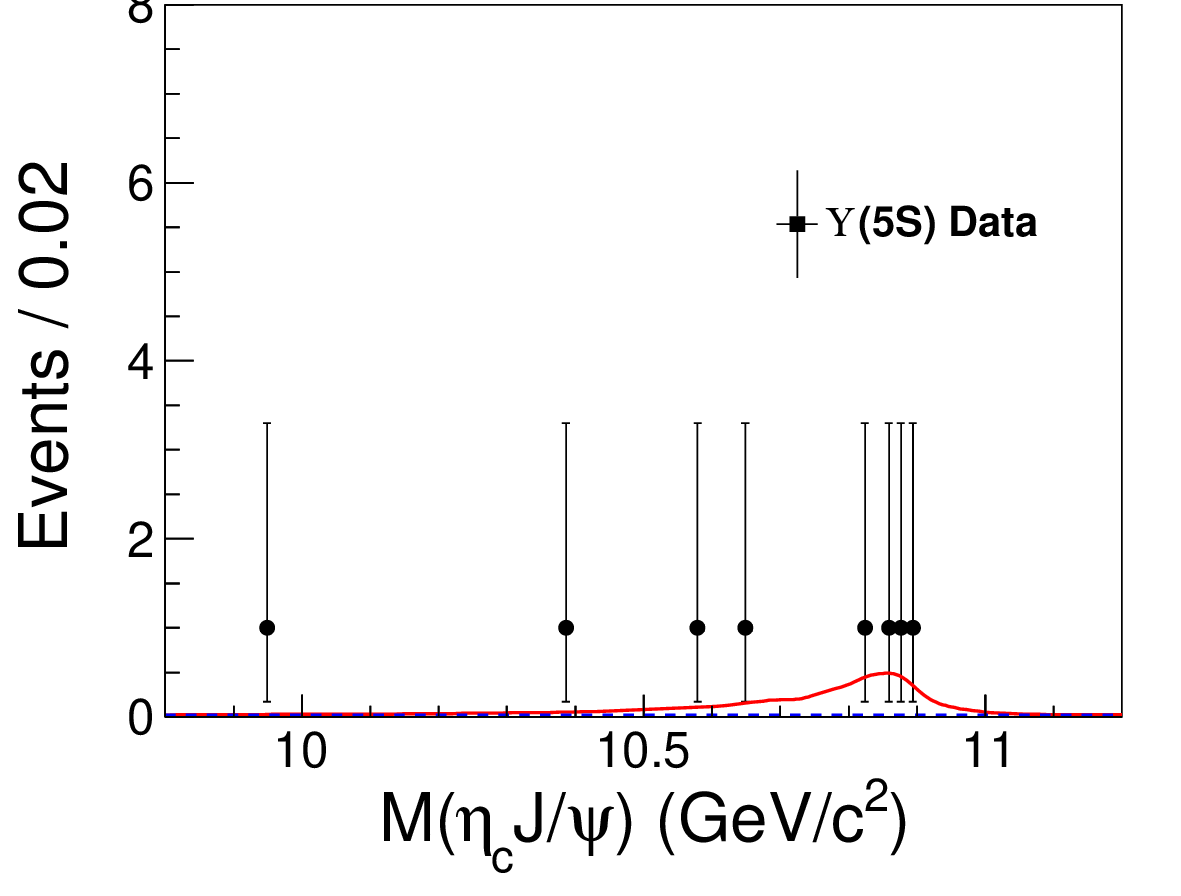}
\end{center}
\caption{Fit to the invariant mass of $\etac\jpsi$ at selected energy points. Dots with error bars show the experimental data, red curves the fit results, and blue dashed curves the background.}\label{fig:MYcc_excJpsi}
\end{figure*}

In this part, we only analyze the datasets at the $\Upsilon(nS)$ resonances as well as the $\Upsilon(4S)$ off-resonance sample, corresponding to a luminosity of $955~\rm fb^{-1}$.
In double charmonium production at $\Upsilon(nS)$ on-resonance and $\Upsilon(4S)$ off-resonance energies, the square of the missing mass in the exclusive reconstruction is required to be within the interval $[-0.05,~0.08]~{\rm GeV}^2/c^4$ to increase the signal purity.
The invariant mass distributions of $\etac\jpsi$ in exclusive reconstruction are shown in Fig.~\ref{fig:MYcc_excJpsi} for data from the selected energy points. 
The  main background are due to combinatorial $\etac$ and $\jpsi$ candidates.
From a study of sideband events in the data, no peaking background is expected.
Unbinned extended maximum likelihood fits are performed to the $\etac\jpsi$ invariant mass distributions except for the $\Upsilon(3S)$ dataset.
Signal components are described using shapes derived from MC simulation, and smoothed using kernel estimation~\cite{rookeyspdf}.
The background components are described with a first order polynomial.
Solely for the $\Upsilon(3S)$ dataset, the upper limit at the 90\% confidence level (C.L.) on the signal yield is estimated~\cite{trolke}.
Cross sections are calculated with the formula
\begin{equation}
    \sigma = \frac{N_{\rm sig}}{\epsilon \mathcal{L} \mathcal{B}(\jpsi\to\LL)\mathcal{B}(\etac\to\rm 6~ channels)},
\label{eq:xsection}
\end{equation}
where $N_{\rm sig}$ is the number of signal events, $\epsilon$ is the reconstruction efficiency, $\mathcal{L}$ is the integrated luminosity, and $\mathcal{B}$ includes the corresponding branching fractions.
The calculated cross sections are shown in Table~\ref{tab:xs_excJpsi}.

\begin{table*}[htp]
\caption{Signal yields and the measured cross sections for different channels at selected c.m. energy points. Here $\mathcal{L}$ is the integrated luminosity; $\epsilon$ is the reconstruction efficiency; $N$ and $\sigma$ are, respectively, the signal yields and measured cross section, where the superscript ``$\rm exc/inc$'' indicates the exclusive/inclusive analysis; and $\sigma^{\rm comb}$ is the cross section combining the two reconstruction analyses. The first uncertainties in $\sigma^{\rm exc/inc}$ are statistical and the second are systematic. The $\sigma^{\rm comb}$ total uncertainties are given including the statistical and systematic.}
\renewcommand\arraystretch{1.2}
\centering
\resizebox{\textwidth}{23mm}{
\begin{tabular}{c   c   c  c  c c c  }
\hline
\hline
		 & $\Upsilon(1S)$	& $\Upsilon(2S)$	&  $\Upsilon(3S)$ &  10.52 GeV	 &   $\Upsilon(4S)$ &   $\Upsilon(5S)$	   \\
\hline
$\mathcal{L}$ [$\rm fb^{-1}]$ & $5.7$  &   $24.9$  &  $2.9$  & $89.4$   & $711.0$	 & $121.4$  \\
$N^{\rm exc}$ & 	 $0.7^{+1.5}_{-0.9}$     & $6.2^{+3.1}_{-2.3}$ & $<1.9$	 & $2.6^{+3.5}_{-2.5}$	 & $45.0^{+8.9}_{-8.2}$	&	 $6.5^{+3.4}_{-2.7}$\\
$\epsilon^{\rm exc}$ & 8.3\%  & 6.9\% & 5.7\% & 5.6\% & 5.6\% & 5.4\%   \\
$\sigma^{\rm exc}~[\rm fb]$  & $57^{+122}_{-73}\pm6$ & $140^{+70}_{-52}\pm14$& $<442$	& $20^{+27}_{-19}\pm6$  & $44^{+9}_{-8}\pm5$	&  $39^{+20}_{-14}\pm7$ \\
\hline
$N^{\rm inc}$ & $23.7\pm12.3$ & $62.0\pm17.9$ & $8.5\pm5.2$  & $94.7\pm23.8$&  $1116.2\pm62.9$& $91.1\pm21.5$ \\
$\epsilon^{\rm inc}$   & 38.6\% & 29.6\%  & 26.4\% & 26.1\% & 25.4\% & 24.7\% \\
$\sigma^{\rm inc}~[\rm fb]$  & $ 89.1 \pm 46.2 \pm 20.5 $ &$ 70.1 \pm 20.2 \pm 8.9 $& $ 91.8 \pm 56.2 \pm 52.3 $  & $ 33.8 \pm 8.5 \pm 2.8 $ 	&$ 52.1 \pm 2.9 \pm 5.0 $ & $ 25.4 \pm 6.0 \pm 2.8 $  \\
\hline
$\sigma^{\rm comb}~[\rm fb]$ &$78.3^{+47.5}_{-43.0}$ & $80.2\pm20.4$ & $87.0^{+71.0}_{-59.0}$ &  $32.5\pm8.5$ & $50.2\pm5.0$ & $27.5\pm6.1$\\
\hline
\end{tabular}
}
\label{tab:xs_excJpsi}
\end{table*}%

\begin{figure*}[htbp]
\begin{center}
    \includegraphics[width=0.30\textwidth]{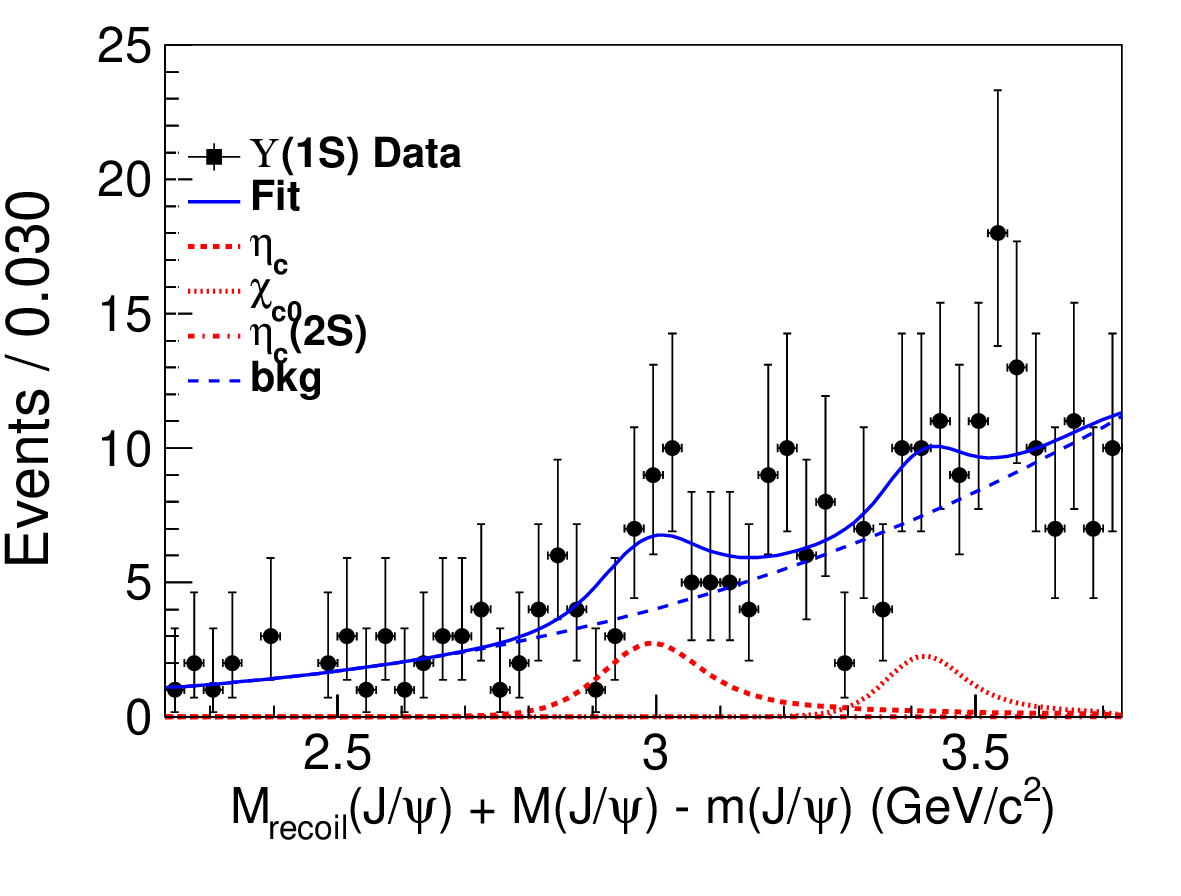}
    \includegraphics[width=0.30\textwidth]{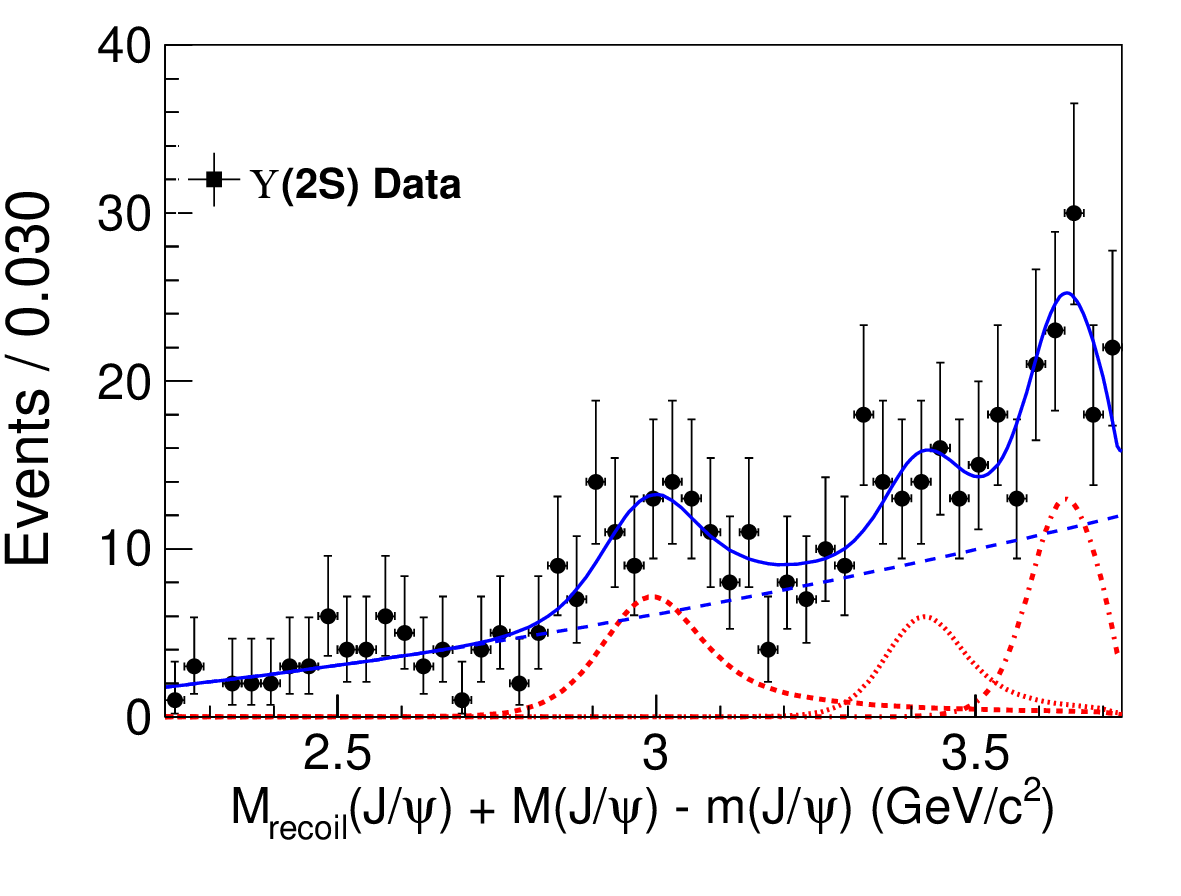}
    \includegraphics[width=0.30\textwidth]{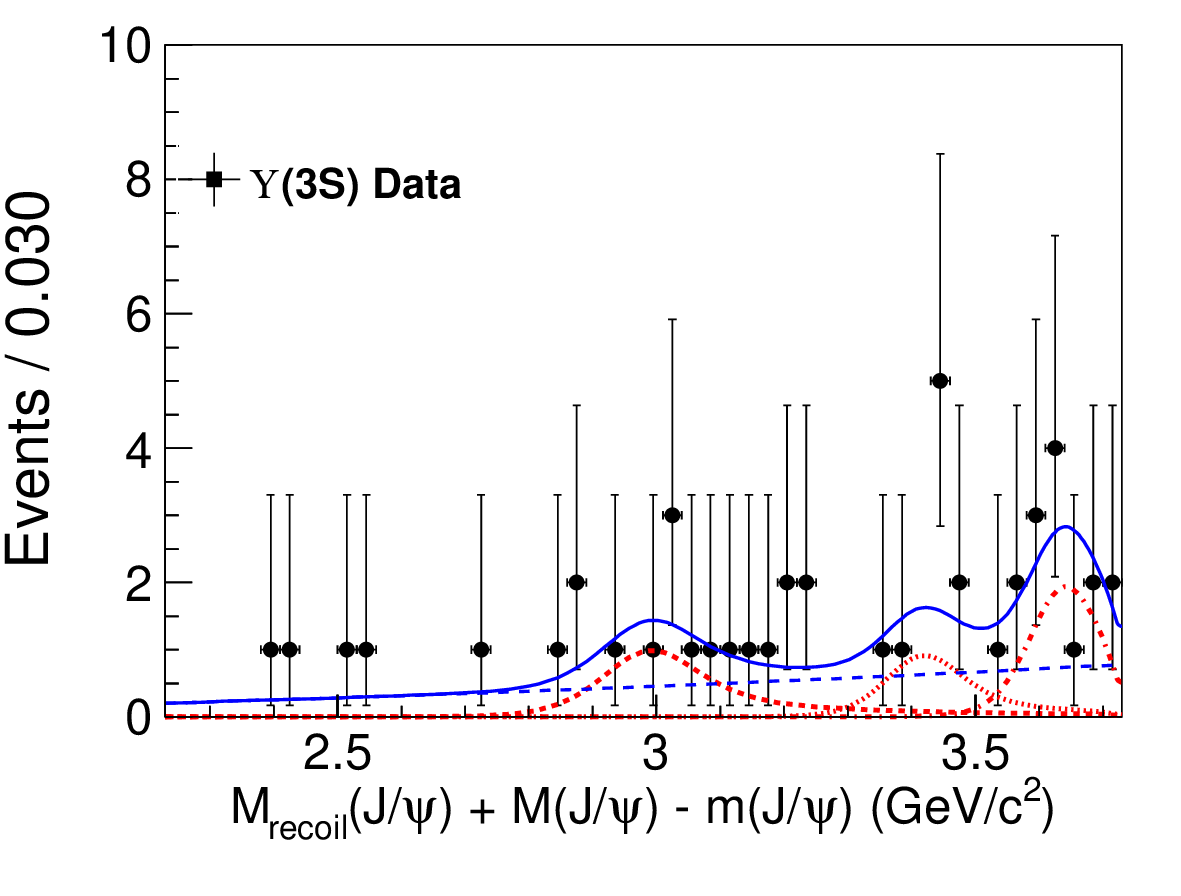}
    \includegraphics[width=0.30\textwidth]{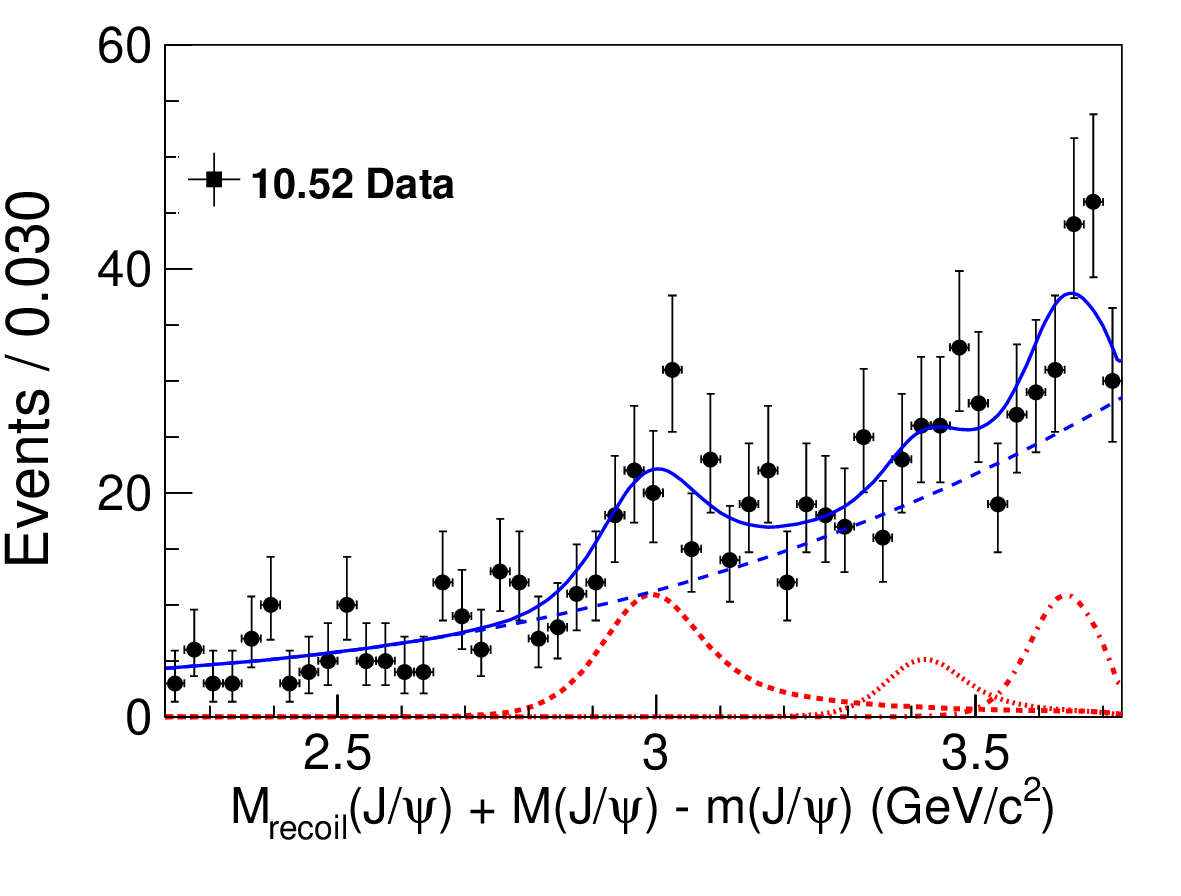}
    \includegraphics[width=0.30\textwidth]{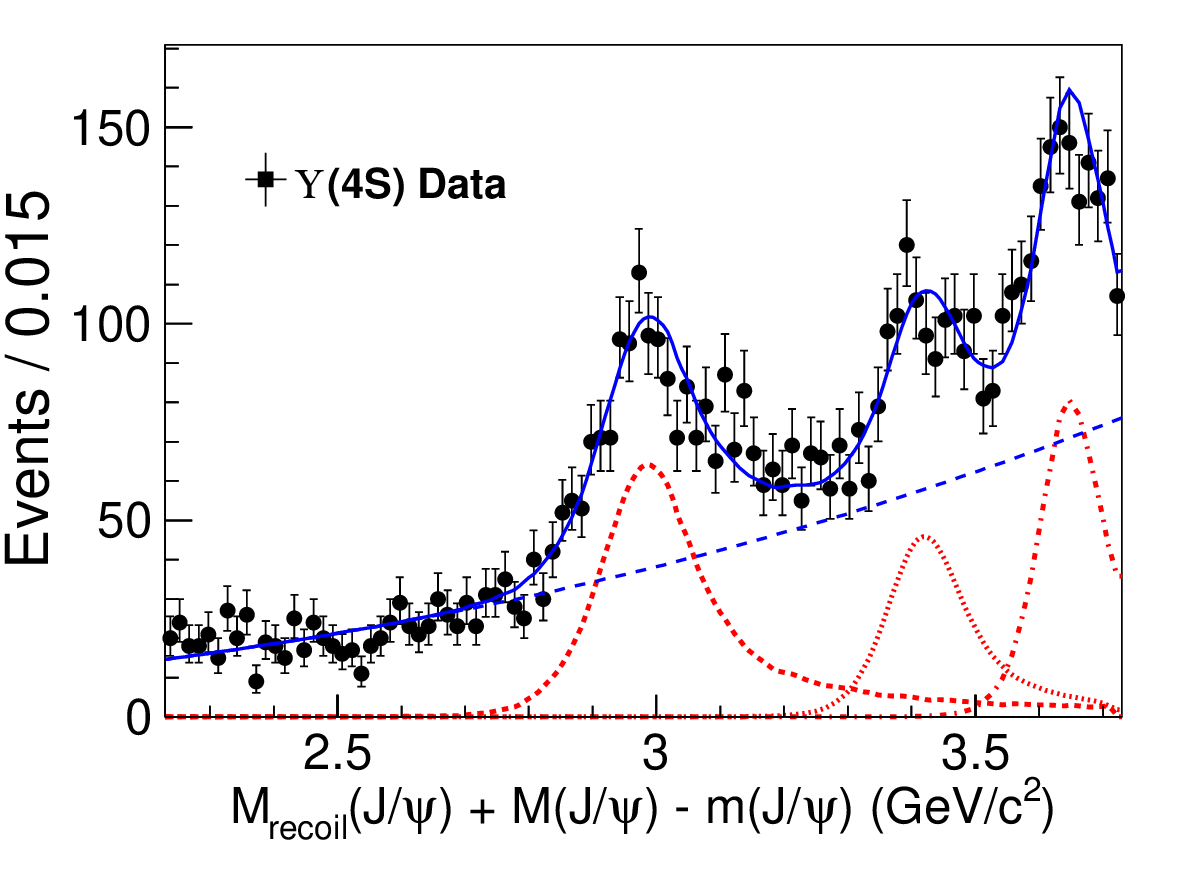}
    \includegraphics[width=0.30\textwidth]{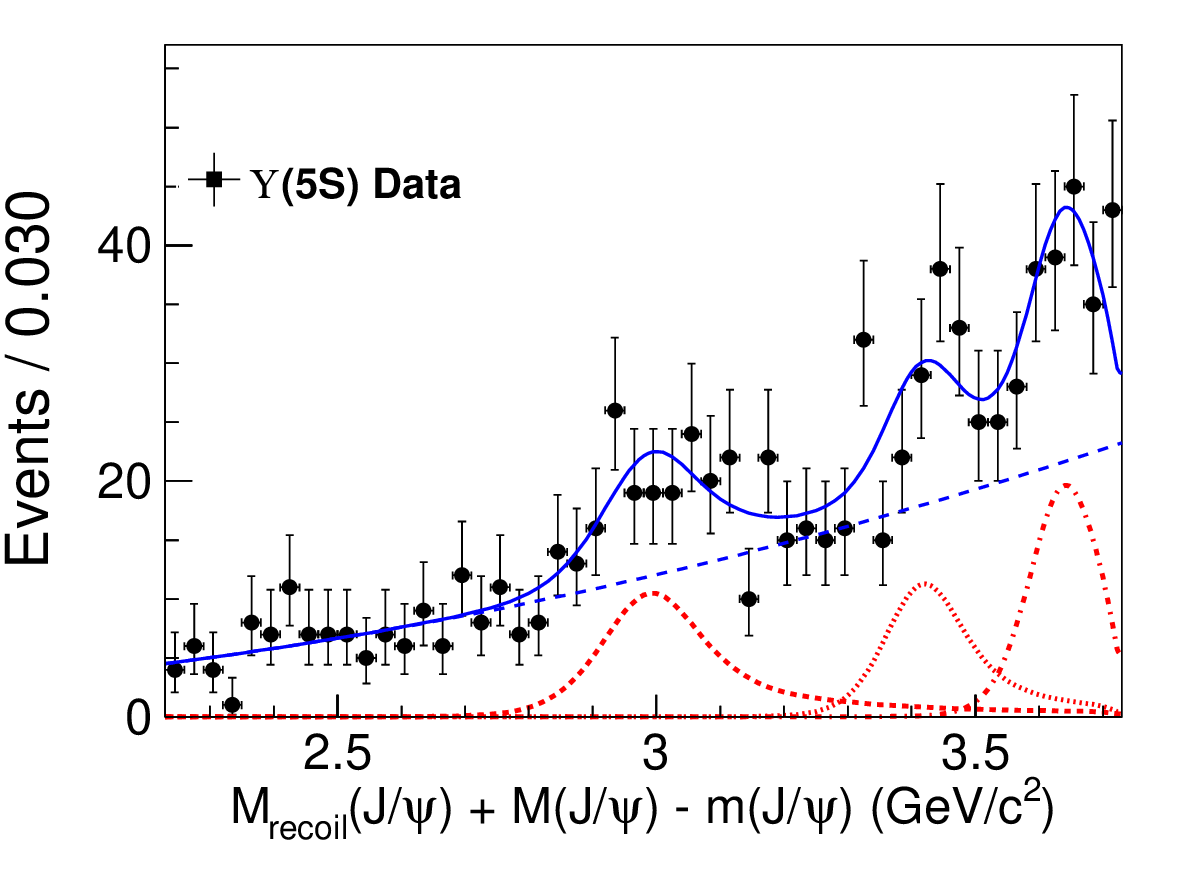}
\end{center}
\caption{Fit to the mass recoiling against $\jpsi$ at the selected energy points. Dots with error bars show the experimental data, blue curves the fit results, and blue long dashed curves the background. The red dashed, dotted, and dashed-dotted curves show the signal for $\eta_c$, $\chi_{c0}$, and $\eta_c(2S)$, respectively.}\label{fig:FitMrJpsi_incJpsi}
\end{figure*}

The $\jpsi$ recoil-mass distributions in the inclusive reconstruction are shown in Fig.~\ref{fig:FitMrJpsi_incJpsi}, where the $\jpsi$ recoil mass is defined as $M_{\rm recoil}(\jpsi)\equiv\sqrt{|p_{\EE}-p_{\jpsi}|^2}/c$ and $p$ is the four-momentum.
To improve the resolution on the recoil mass, we replace the recoil mass with $M_{\rm recoil}(\jpsi)+M(\jpsi)-m(\jpsi)$, where $M(\jpsi)$ is the reconstructed $\jpsi$ mass and $m(\jpsi)$ is the nominal mass from Ref.~\cite{PDG}.
Clear $\etac,~\chi_{c0}$, and $~\etac(2S)$ signals are found in the recoil-mass spectra, as in previous Belle measurements~\cite{BelleCC}~\cite{BelleCC2}.
Unbinned extended maximum likelihood fits are performed to the recoil-mass spectra.
The signals are described using shapes derived from MC simulation smeared with a Gaussian.
Parameters of the Gaussian functions are free for the fit to the $\Upsilon(4S)$ on-resonance data sample, but fixed to the values obtained in the $\Upsilon(4S)$ fit result when fitting to other energy points.
Background is described with a third-order-polynomial.
Signal yields from the fits are listed in Table~\ref{tab:xs_excJpsi} together with the cross sections.
The cross sections here are calculated using a formula similar to Eq.~\ref{eq:xsection} but omitting the $\etac$ branching fractions.

\begin{figure}[htbp]
\begin{center}
    \includegraphics[width=0.45\textwidth]{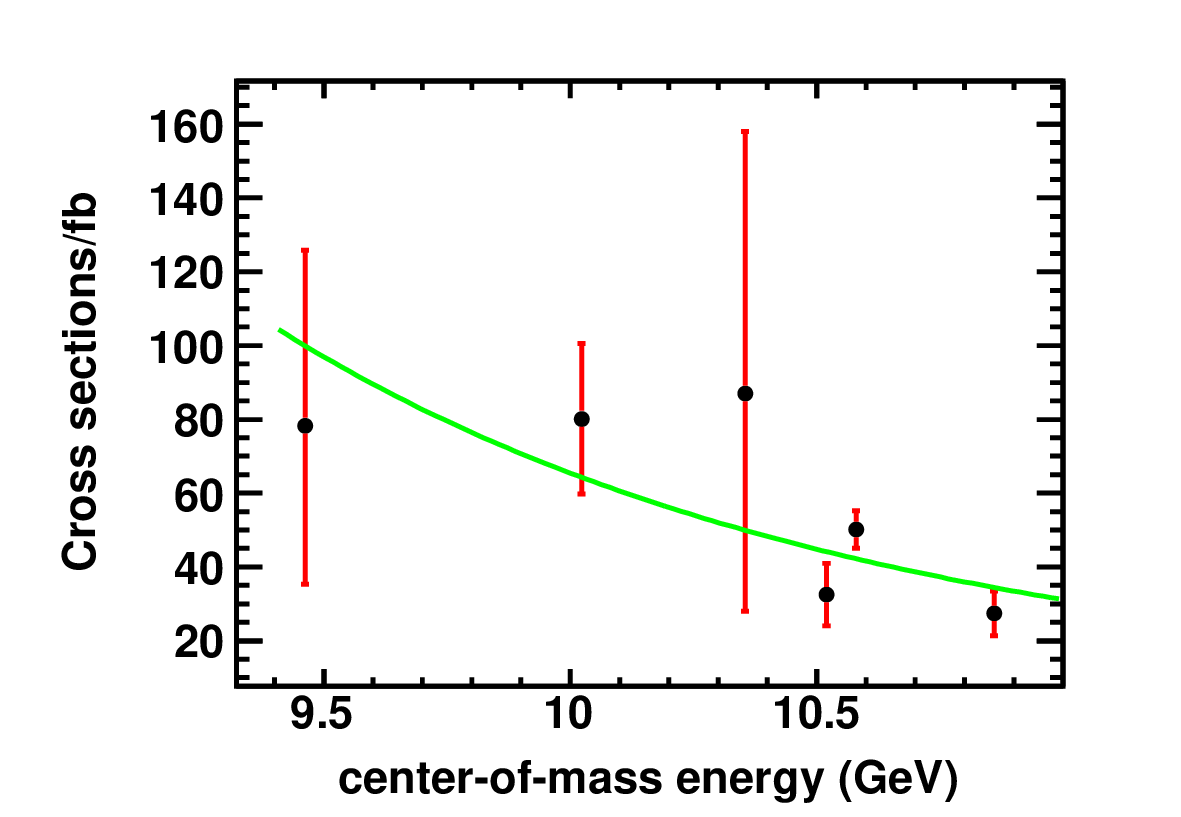}
\end{center}
\caption{Combined results of the measured cross sections of $\EE\to\etac\jpsi$ at the $\Upsilon(nS)$ on-resonance and $\Upsilon(4S)$ off-resonance energy points. The green curve is the fit result using Eq.~\ref{eq:xsfunction}.}\label{fig:xs_YnS2etacJpsi}
\end{figure}

We combine cross sections measured from the two reconstruction methods with the following approach. 
First, we extract the cross-section-dependent likelihood distributions from the two methods.
For a given cross section in the certain data sample, the possibility that we observe the current number of signal events is estimated according to the fits.
(For the exclusive reconstruction with $\Upsilon(3S)$ data, we assume the number of the observed events follows the Poisson distribution.)
Consequently, a cross-section-dependent joint probability density function (PDF) is obtained.
We smear the PDF with Gaussian functions whose widths model the systematic uncertainties that are discussed below.
The peak of the PDF is taken as the nominal result, and the positions bounding 68\% of the total integrated area under the PDF are taken as the uncertainties.
The final cross section results are listed in Table~\ref{tab:xs_excJpsi} and plotted in Fig.~\ref{fig:xs_YnS2etacJpsi}.

We fit the cross sections to extrapolate to the $\etac\jpsi$ threshold region to estimate the continuum contribution if any signal is found there.
There should be two sources of $\etac\jpsi$ production:
one is $\EE\to\gamma^*\to\etac\jpsi$, which is so called continuum production, 
and the other is $\EE\to\Upsilon(nS)\to\gamma^*\to\etac\jpsi$.
This mechanism should be similar to the production of $\EE\to\MM$.
Thus, we estimate the fractions produced from the continuum 
relative to $\EE \to \MM$ total cross sections at $\Upsilon(nS)$ energy point in Ref.~\cite{12Smm},
where the corresponding fractions of the continuum production are about $5/6$ and $4.5/4.75$ for $\Upsilon(1S)$ and $\Upsilon(2S)$, respectively. 
Since we do not find a signal in the $\Upsilon(3S)$ dataset, and the uncertainty at this energy point is very large, we use a fraction of one at this energy point.
For the $\Upsilon(4S)$ and $\Upsilon(5S)$, we assume that signals originate exclusively from continuum production.
We fit the cross sections of $\EE\to\etac\jpsi$ from the continuum production at $\Upsilon(1/2S)$ along with the $\EE\to\etac\jpsi$ at other energy points with the empirical  function
\begin{equation}
\sigma  =  A \frac{ \sqrt{2\mu\Delta M} } { (\frac{s}{s_0})^{n} }, 
\label{eq:xsfunction}
\end{equation}
where $\mu = \frac{m(\etac)m(\jpsi)}{m(\etac)+m(\jpsi)}$ is the reduced mass,$~\Delta M  = \sqrt{s} - m(\etac) - m(\jpsi)$ is the mass difference, $m(\etac)$ and $m(\jpsi)$ are the $\etac$ and $\jpsi$ nominal masses~\cite{PDG}; $s_0\equiv(10.58~\gevcc)^2$ is the c.m. energy at the $\Upsilon(4S)$ resonance, $A$ and $n$ are free parameters whose best-fit values are $A=(11.4\pm0.9)~c^2\cdot\rm fb/GeV$, and $n = 4.5\pm1.3$.

\begin{figure*}[hbtb]
\begin{center}
    \includegraphics[width=0.99\textwidth]{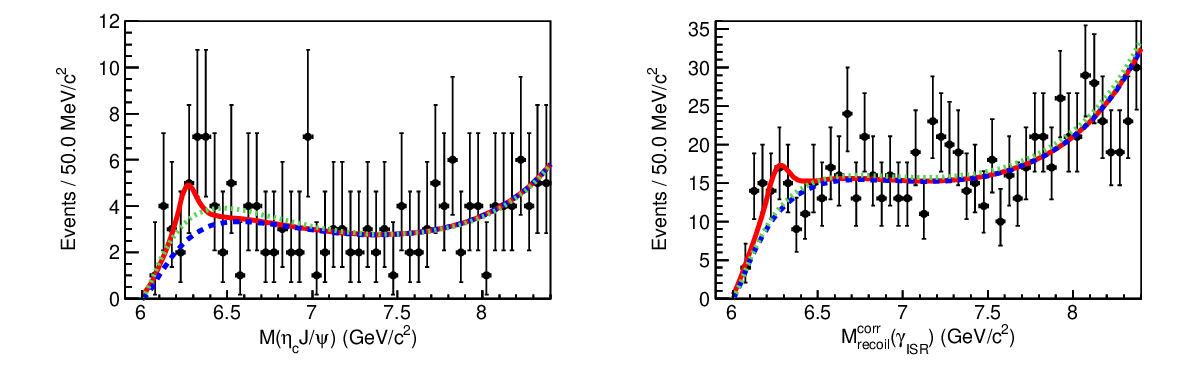}
\end{center}
\caption{Simultaneous fit result to the invariant mass of $\etac\jpsi$ (left) and the $\gamma$ recoil mass (right). In each panel, dots with error bars are from data, the red solid curve is the best fit result, the blue dashed curve the background component from the best fit, and the green dotted curve is the fit result without the signal components.}\label{fig:SimFit_Ycc}
\end{figure*}

\begin{figure*}[hbtb]
\begin{center}
    \includegraphics[width=0.32\textwidth]{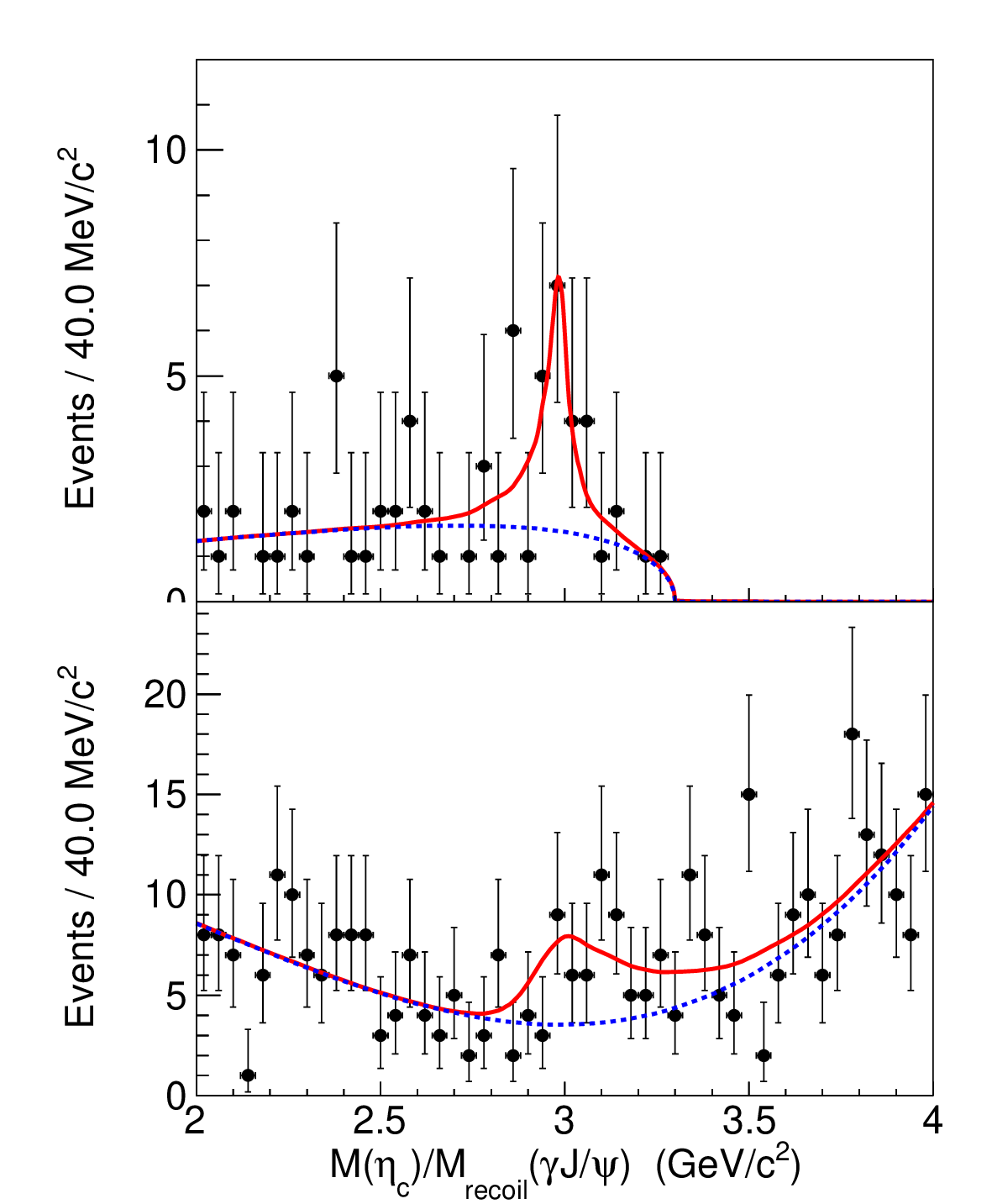}
    \includegraphics[width=0.32\textwidth]{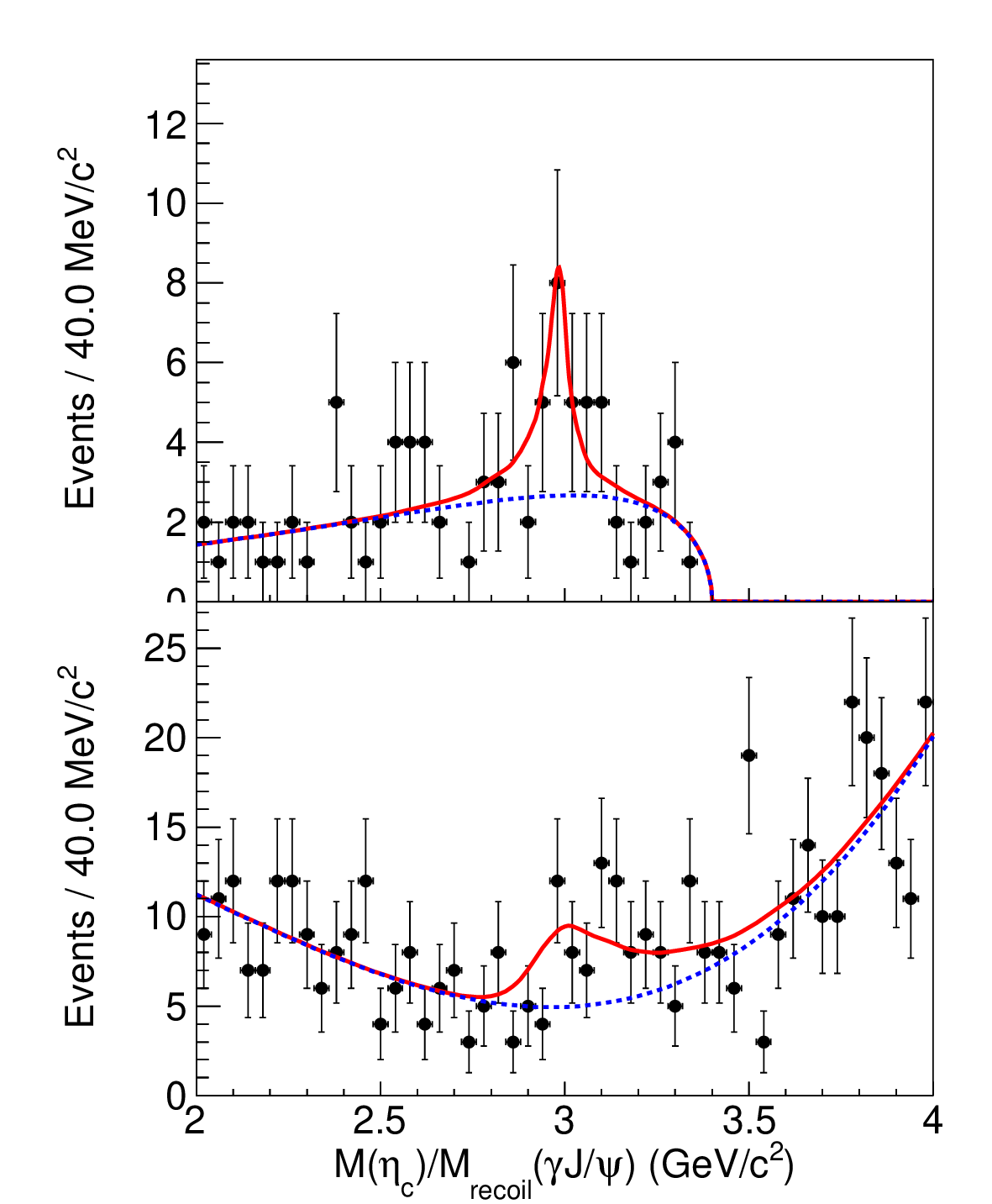}
    \includegraphics[width=0.32\textwidth]{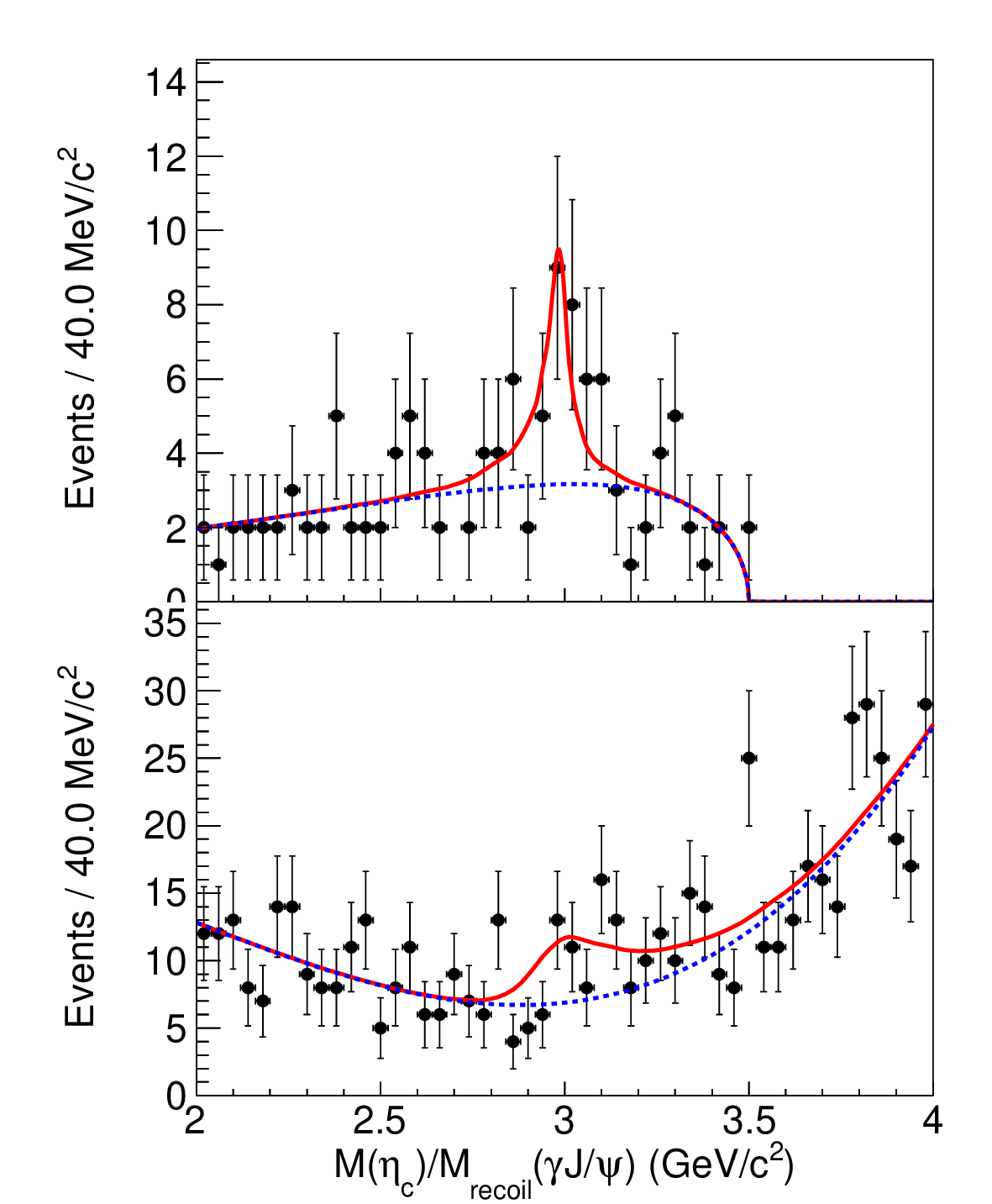}
\end{center}
\caption{Simultaneous fit result to the invariant mass of reconstructed $\etac$ (top) and the $\gisr\jpsi$ recoil mass (bottom). From left to right are events with  $M(\etac\jpsi)$ and $M_{\rm recoil}(\gamma)\in[6.0,~6.4],~[6.0,~6.5]$, and $[6.0,~6.6]~\gevcc$. Dots with error bars are from data, the red solid curve is the best fit result, and the blue dashed curve is the background component from the best fit.}\label{fig:SimFit_Etac}
\end{figure*}

\begin{figure}[hbtp]
\begin{center}
    \includegraphics[width=0.45\textwidth]{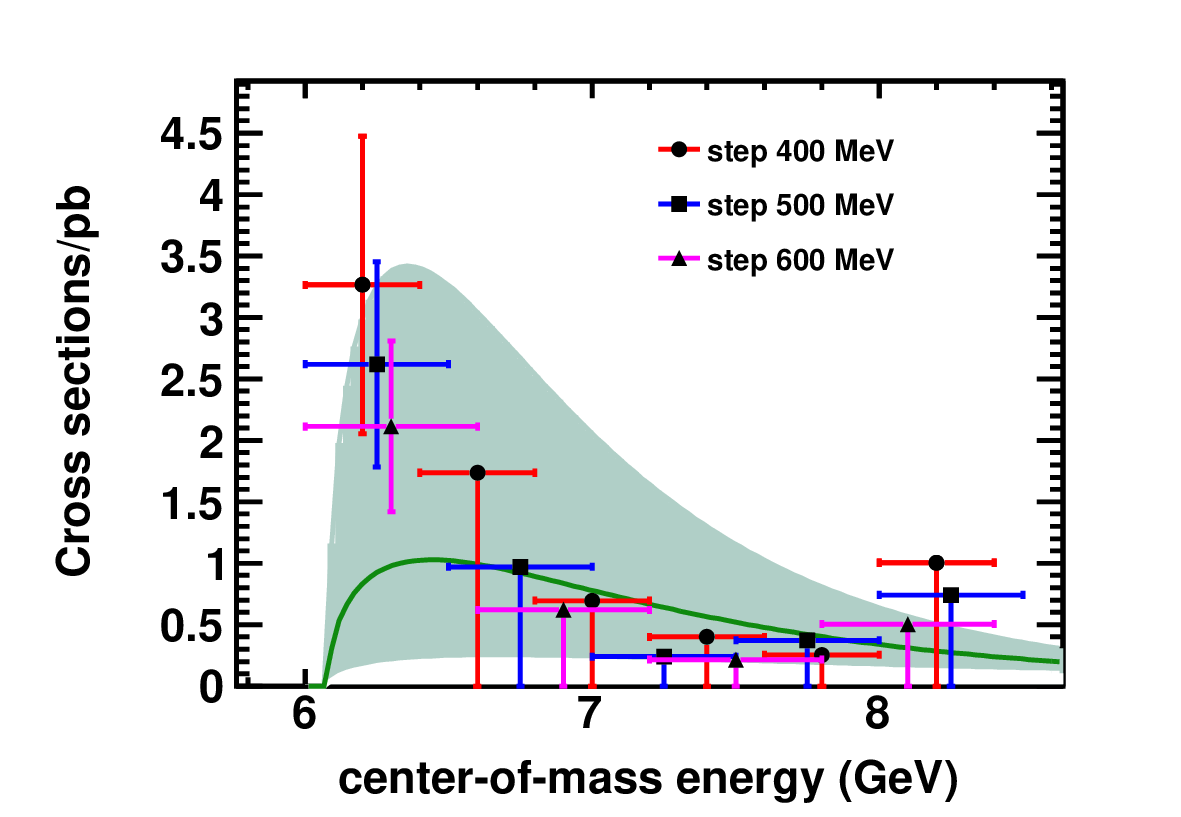}
\end{center}
\caption{Measured cross sections of $\EE\to\etac\jpsi$ near the threshold. From left to right are  the cross sections measured in different step sizes (0.4, 0.5, 0.6 $\gevcc$). Green solid curve is the extrapolation from the fit to near-resonance energy points. The shadow area covers the $\pm1\sigma$ range of the extrapolations.
}\label{fig:xs_Ycc2etacJpsi}
\end{figure}

\begin{table}[hbpb]
\caption{Measured cross sections for different $M(\etac\jpsi)$ and $M^{\rm corr}_{\rm recoil}(\gisr)$ regions. Here, $N_{\rm prod}$ is the number of produced $\etac\jpsi$ signal events, and $\sigma$ is the calculated cross section.}
\begin{center}
\begin{tabular}{c  c  c }
\hline
\hline
	regions ($\gevcc$)	&   $N_{\rm prod}~[\times10^{2}]$  & $\sigma~[\rm pb]$  	\\
\hline
	 $[6.0,~6.4]$		& $13.1\pm3.6$		 &  $3.3\pm0.9\pm0.8$		 \\
	 $[6.4,~6.8]$		& $<8.2$	         &  $<1.7$		 \\
	 $[6.8,~7.2]$		& $<3.9$			 &  $<0.7$		 \\
	 $[7.2,~7.6]$		& $<2.7$			 &  $<0.4$		 \\
	 $[7.6,~8.0]$		& $<2.1$			 &  $<0.3$		 \\
	 $[8.0,~8.4]$		& $<10.4$	 		 &  $<1.0$		 \\
\hline

	 $[6.0,~6.5]$		& $13.4\pm4.0$ 	 &  $2.7\pm0.8\pm0.2$		 \\
	 $[6.5,~7.0]$		& $<6.1$			 &  $<1.0$		 \\
	 $[7.0,~7.5]$		& $<1.9$	 		 &  $<0.2$	 \\
	 $[7.5,~8.0]$		& $<3.8$			 &  $<0.4$		 \\
	 $[8.0,~8.5]$		& $<9.9$			 &  $<0.7$		 \\

\hline

	$[6.0,~6.6]$		& $13.3\pm4.2$	     &  $2.1\pm0.7\pm0.2$		 \\
	$[6.6,~7.2]$		& $<5.0$			 &  $<0.6$		 \\
	$[7.2,~7.8]$		& $<2.3$	 		 &  $<0.2$	 \\
	$[7.8,~8.4]$		& $<7.4$			 &  $<0.5$	 \\
	
\hline
\end{tabular}
\end{center}
\label{tab:xs_Ycc2etacJpsi}
\end{table}%

\subsection{$\EE\to\etac\jpsi$ near threshold}

In this part, we use the entire $980~\rm fb^{-1}$ Belle dataset.
For exclusive reconstruction of $\EE\to\etac\jpsi$ near-threshold events, the squared recoil mass of the $\etac\jpsi$ system ($|p_{\EE}-p_{\etac\jpsi}|^2/c^2$) is required to be within the interval $[-1,~2]~{\rm GeV}^2/c^4$ to improve the signal purity. 
Here, the mass window is larger than in the previous section because of the possibility of a second ISR photon.
To suppress the possible background from $\Upsilon(4S)\to B \bar B$, the ratio of the second to the zeroth order Fox-Wolfram moments~\cite{fwmomentum} is required to be greater than 0.13.
To improve the resolution of the $\Xcc$ signal in the inclusive reconstruction, we use the corrected recoil mass $M^{\rm corr}_{\rm recoil}(\gisr)\equiv M_{\rm recoil}(\gisr)-M_{\rm recoil}(\gisr\jpsi)+m(\etac)$, 
where $M_{\rm recoil}(\gisr)$ and $M_{\rm recoil}(\gisr\jpsi)$ are the recoil masses of $\gisr$ and $\gisr\jpsi$, respectively, and $m(\etac)$ is the $\etac$ nominal mass~\cite{PDG}.

The invariant mass of $\etac\jpsi$ and the recoil mass of $\gamma_{\rm ISR}$ are shown in Fig.~\ref{fig:SimFit_Ycc}.
Events in common between the exclusive and inclusive samples are removed from the inclusive reconstruction to avoid double counting.
The number of events increases near threshold in the mass spectrum of $\etac\jpsi$, but no similar enhancement is seen in the recoil mass of $\gisr$.
A simultaneous unbinned maximum likelihood fit for the $\etac\jpsi$ invariant mass and $\gisr$ recoil mass is performed.
The signal-yield fractions from the two reconstruction methods are fixed to the corresponding branching fractions and reconstruction efficiencies.
The background shapes are parameterized with the ARGUS function, whose parameters are obtained from the fit to the $\etac$ and $\jpsi$ sideband events.
The signals are described with a Breit-Wigner function with free mass and width convolved with the Gaussian functions from the resolution study.
The fit results are shown in Fig.~\ref{fig:SimFit_Ycc}.
The significance of the Breit-Wigner peak component is $2.1\sigma$, with mass and width of $(6267\pm43)~\mevcc$ and $(121\pm72)~\rm MeV$, respectively.
The signal yields are $9\pm4$ and $23\pm11$ from the exclusive and inclusive methods, respectively.

The reconstructed $\etac$ mass and $\gisr\jpsi$ recoil mass spectra for events satisfying $M(\etac\jpsi)$ and $M_{\rm recoil}(\gamma)\in[6.0,~6.4~\rm or~6.5~\rm or~6.6]~\gevcc$ are shown in Fig.~\ref{fig:SimFit_Etac}.
Fit results to other mass regions are shown in the Appendix. 
No peaking backgrounds are expected according to the study of the generic MC sample.
We perform a simultaneous unbinned maximum likelihood fit to the reconstructed $\etac$ mass and $\gamma_{\rm ISR}\jpsi$ recoil mass.
The signal-yield fractions for the two reconstruction methods are fixed to the products of the corresponding branching fractions and reconstruction efficiencies obtained from simulated signal events. 
The signal shapes are described using shapes derived from MC simulation and smoothed using 
kernel estimation~\cite{rookeyspdf}, while the background is described with an ARGUS function~\cite{ARGUS} and third-order-polynomial for the exclusive and inclusive analyses, respectively.

The fit results are shown in Fig.~\ref{fig:SimFit_Etac}.
The significances of the $\etac$ components are $3.9,~3.3$, and $3.5\sigma$ for events from the three mass regions, respectively.
The invariant mass of $\etac$ and $\gisr\jpsi$ recoil mass for the other $M(\etac\jpsi)/M_{\rm recoil}(\gisr)$ mass regions are shown in the Appendix, as well as the fit results.
No evident signals are found in those distributions, and the upper limits of the number of produced events in different $\etac\jpsi$ mass regions are estimated at 90\% C.L. with the following method.
First, the profile likelihood distribution as a function of the number of produced events is extracted from the fit. 
Then this likelihood distribution is smeared with a Gaussian function whose width models the systematic uncertainty.
This smeared distribution is then integrated from zero to infinity.
The point at which the integral reaches 90\% of the total value is taken as the upper limit.
The number of produced events and the upper limits from different $M(\etac\jpsi)$ and $M_{\rm recoil}(\gisr)$ mass regions are listed in Table~\ref{tab:xs_Ycc2etacJpsi}.


\begin{table}[hbtp]
\caption{Summary of the systematic uncertainties (in \%), excluding those for the fitting procedures. 
For both photon detection and total uncertainties, the main number shows the uncertainty for the $\Xcc$ search, and the parentheses show the uncertainty for the cross section measurements at the selected on- (off-)resonance points.
}
\begin{center}
\begin{tabular}{c  c  c}
\hline
\hline
	source		&  exclusive reconstruction & inclusive reconstruction 	\\
\hline
    Tracking		     &  1.4                 &   0.7        \\
    Photon detection	 &  0.0                 &   2.0 (0.0)         \\
    PID                  & 9.2                 &   7.2         \\
    $K_S$ selection      &  0.3                 &   0.0         \\
    $\pi^0$ selection &3.5                 &   0.0         \\
    $\etac$ decays       &  0.9                 &   0.0         \\
\hline
    $\jpsi$ decays & \multicolumn{2}{c}{0.5}  \\
    Luminosity     &    \multicolumn{2}{c}{1.4}  \\
    Generator	&  \multicolumn{2}{c}{1.0}  \\
\hline
	Sum			     & \multicolumn{2}{c}{8.1 (7.8)} 	\\	
\hline
\end{tabular}
\end{center}
\label{tab:sysSum}
\end{table}%

The effective luminosity in each mass region is calculated according to Ref.~\cite{effLumi}. 
Using this, we calculate the cross sections near $\etac\jpsi$ mass threshold with an equation analogous to Eq.~\ref{eq:xsection}; these are plotted in Fig.~\ref{fig:xs_Ycc2etacJpsi} as points with errors.
We extrapolate the lineshape of the measured cross sections near $\Upsilon(nS)$ resonances according to Eq.~\ref{eq:xsfunction}, and plot it in Fig.~\ref{fig:xs_Ycc2etacJpsi} as the solid curve for comparison with the measurements. 
We vary the parameters of the extrapolations based on the uncertainties and correlations. The range of the extrapolations are also shown on the plot.
The measured cross sections near $\etac\jpsi$ mass threshold are consistent with the extrapolations from the $\Upsilon(nS)$ energy region according to their uncertainty.

\section{Systematic uncertainty}

Possible sources of systematic uncertainty include tracking, ISR photon detection, PID, $K_S^{0}$ reconstruction, $\pi^0$ reconstruction, the fitting procedure, integrated luminosity, and the $\etac$ and $\jpsi$ branching fractions, as listed in Table~\ref{tab:sysSum}.

The difference in tracking efficiency for tracks with momenta above $200~\mevcc$ between data and MC is $(-0.13\pm0.30\pm0.10)\%$ per track.
The uncertainty of ISR photon detection efficiency, studied using radiative bhabha events, is 2.0\%. 
This is only taken into account in the $\Xcc$ search. 
We apply a reconstruction uncertainty of $0.35\%$ per track in our analysis estimated by using partially reconstructed $D^{*+}\to D^0\pi^+,~D^0\to\pipi K_S^0$, and $K^0_S\to\pipi$ events.
According to the measurement of PID efficiency using the control sample $D^{*}\to D^0\pi$ with $D^0\to K^-\pi^+$, we assign uncertainties of 1.1\% for each kaon and 0.9\% for each pion.
For $K_S^0$ selection, we take 2.2\% as the systematic uncertainty, following Ref.~\cite{kserr}.
For $\pi^0$ selection, the uncertainty is 2.3\% according to a study of the $\tau\to\pi\pi^0\nu_{\tau}$ control sample~\cite{pi0err}.

Since we are using multiple channels to reconstruct $\etac$, the uncertainties of reconstruction efficiencies from these channels are combined using
\begin{equation}
\delta = \frac{\sum_{i}\delta_{i}\epsilon_{i}\mathcal{B}_{i}}{\sum_{i}\epsilon_{i}\mathcal{B}_{i}},
\label{eq:sys}
\end{equation}
where $\delta_{i}$, $\epsilon_{i}$, and $\mathcal{B}_{i}$ are the systematic uncertainty, reconstruction efficiency, and branching fraction from the $i$-th $\etac$ channel, respectively.

For the fitting procedure, we change the fitting range and the background function.
The difference in signal yields from the nominal and alternate fits is taken as the systematic uncertainty, as shown in the Appendix.
For the fits with no significant signals, this systematic uncertainty is included in the upper limit estimation by taking the alternative fit result with the largest upper limit for the signal yield.


The uncertainty of the integrated luminosity is about 1.4\%. 
The uncertainty of the branching fraction of $\jpsi\to\LL$, 0.5\%, is taken from Ref.~\cite{PDG}.
The uncertainty due to the branching fractions of $\etac$ decays is studied with pseudo-experiments, in which we vary the branching fractions of $\etac$ decays within $1\sigma$, and calculate the overall reconstruction efficiency.
After 1,000 trials, we obtain a distribution of reconstruction efficiencies, which is subsequently fit to a Gaussian.
The width of this Gaussian is taken as the systematic uncertainty associated with the $\etac$-decay branching fractions.

We are using the {\sc phokhara} generator to simulate our signal events. The discrepancy of the energy of ISR photons between the simulation and the theoretical calculation~\cite{ISRtheory} is less than 0.1\%, with a statistical uncertainty of less than 1.0\%; thus, we take the uncertainty of the generator {\sc phokhara} to be 1.0\% as a conservative value.

The total systematic uncertainty is obtained by adding the individual
components in quadrature.
We use Eq.~\ref{eq:sys} to combine the systematic uncertainties from the two methods used in the cross section measurements near threshold.

\section{Summary}

We perform the first search for a double-charmonium state with $\EE\to\etac\jpsi$ near threshold via the ISR process.
No significant signal of the double charmonium state is found in several bins of the invariant mass of $\etac\jpsi$ (for exclusive reconstruction) and the recoil mass of $\gisr$ (for inclusive reconstruction).
We measure the $\eta_c\jpsi$ production cross sections in several bins of the invariant mass of $\etac\jpsi$ and the recoil mass of $\gamma_{\rm ISR}$.
The cross sections for $\EE\to\etac\jpsi$ nearest the threshold are significantly larger than in neighboring bins.
Evidence with a statistical significance greater than $3.3\sigma$ is found for double charmonium production near the $\etac\jpsi$ threshold.
The cross sections of double charmonium production at $\Upsilon(nS)$ on-resonance and $\Upsilon(4S)$ off-resonance data samples are also measured.
The cross sections are fitted with a function $\sigma \propto 1/s^n$, and extrapolated to the lower $\etac\jpsi$ mass regions,
where consistency within $1\sigma$ with our measurements in those regions is observed, albeit with relatively large measurement uncertainties.
The search for double charmonium production at Belle II is expected to be revisited as SuperKEKB integrated luminosity reaches a few $\rm ab^{-1}$ or more.

\acknowledgments

This work, based on data collected using the Belle detector, which was
operated until June 2010, was supported by 
the Ministry of Education, Culture, Sports, Science, and
Technology (MEXT) of Japan, the Japan Society for the 
Promotion of Science (JSPS), and the Tau-Lepton Physics 
Research Center of Nagoya University; 
the Australian Research Council including grants
DP210101900, 
DP210102831, 
DE220100462, 
LE210100098, 
LE230100085; 
Austrian Federal Ministry of Education, Science and Research (FWF) and
FWF Austrian Science Fund No.~P~31361-N36;
the National Natural Science Foundation of China under Contracts
No.~11675166,  
No.~11705209;  
No.~11975076;  
No.~12135005;  
No.~12175041;  
No.~12161141008; 
Key Research Program of Frontier Sciences, Chinese Academy of Sciences (CAS), Grant No.~QYZDJ-SSW-SLH011; 
Project ZR2022JQ02 supported by Shandong Provincial Natural Science Foundation;
the Ministry of Education, Youth and Sports of the Czech
Republic under Contract No.~LTT17020;
the Czech Science Foundation Grant No. 22-18469S;
Horizon 2020 ERC Advanced Grant No.~884719 and ERC Starting Grant No.~947006 ``InterLeptons'' (European Union);
the Carl Zeiss Foundation, the Deutsche Forschungsgemeinschaft, the
Excellence Cluster Universe, and the VolkswagenStiftung;
the Department of Atomic Energy (Project Identification No. RTI 4002) and the Department of Science and Technology of India; 
the Istituto Nazionale di Fisica Nucleare of Italy; 
National Research Foundation (NRF) of Korea Grant
Nos.~2016R1\-D1A1B\-02012900, 2018R1\-A2B\-3003643,
2018R1\-A6A1A\-06024970, RS\-2022\-00197659,
2019R1\-I1A3A\-01058933, 2021R1\-A6A1A\-03043957,
2021R1\-F1A\-1060423, 2021R1\-F1A\-1064008, 2022R1\-A2C\-1003993;
Radiation Science Research Institute, Foreign Large-size Research Facility Application Supporting project, the Global Science Experimental Data Hub Center of the Korea Institute of Science and Technology Information and KREONET/GLORIAD;
the Polish Ministry of Science and Higher Education and 
the National Science Center;
the Ministry of Science and Higher Education of the Russian Federation, Agreement 14.W03.31.0026, 
and the HSE University Basic Research Program, Moscow; 
University of Tabuk research grants
S-1440-0321, S-0256-1438, and S-0280-1439 (Saudi Arabia);
the Slovenian Research Agency Grant Nos. J1-9124 and P1-0135;
Ikerbasque, Basque Foundation for Science, Spain;
the Swiss National Science Foundation; 
the Ministry of Education and the Ministry of Science and Technology of Taiwan;
and the United States Department of Energy and the National Science Foundation.
These acknowledgements are not to be interpreted as an endorsement of any
statement made by any of our institutes, funding agencies, governments, or
their representatives.
We thank the KEKB group for the excellent operation of the
accelerator; the KEK cryogenics group for the efficient
operation of the solenoid; and the KEK computer group and the Pacific Northwest National
Laboratory (PNNL) Environmental Molecular Sciences Laboratory (EMSL)
computing group for strong computing support; and the National
Institute of Informatics, and Science Information NETwork 6 (SINET6) for
valuable network support.
E. Won is partially
supported by the NRF grant 2022R1A2B5B02001535 and
J. H. Yin and E. Won are by 2019H1D3A1A01101787.


\appendix
\section{Appendix}
\subsection{Fit plots for $\EE\to\etac\jpsi$ near threshold }

Here we provide the fitting plots for different $M(\etac\jpsi)$ mass regions.

\subsubsection{Step size 400 $\mevcc$}


\begin{figure}[htbp]
\begin{center}
\flushleft
    \includegraphics[width=0.29\textwidth]{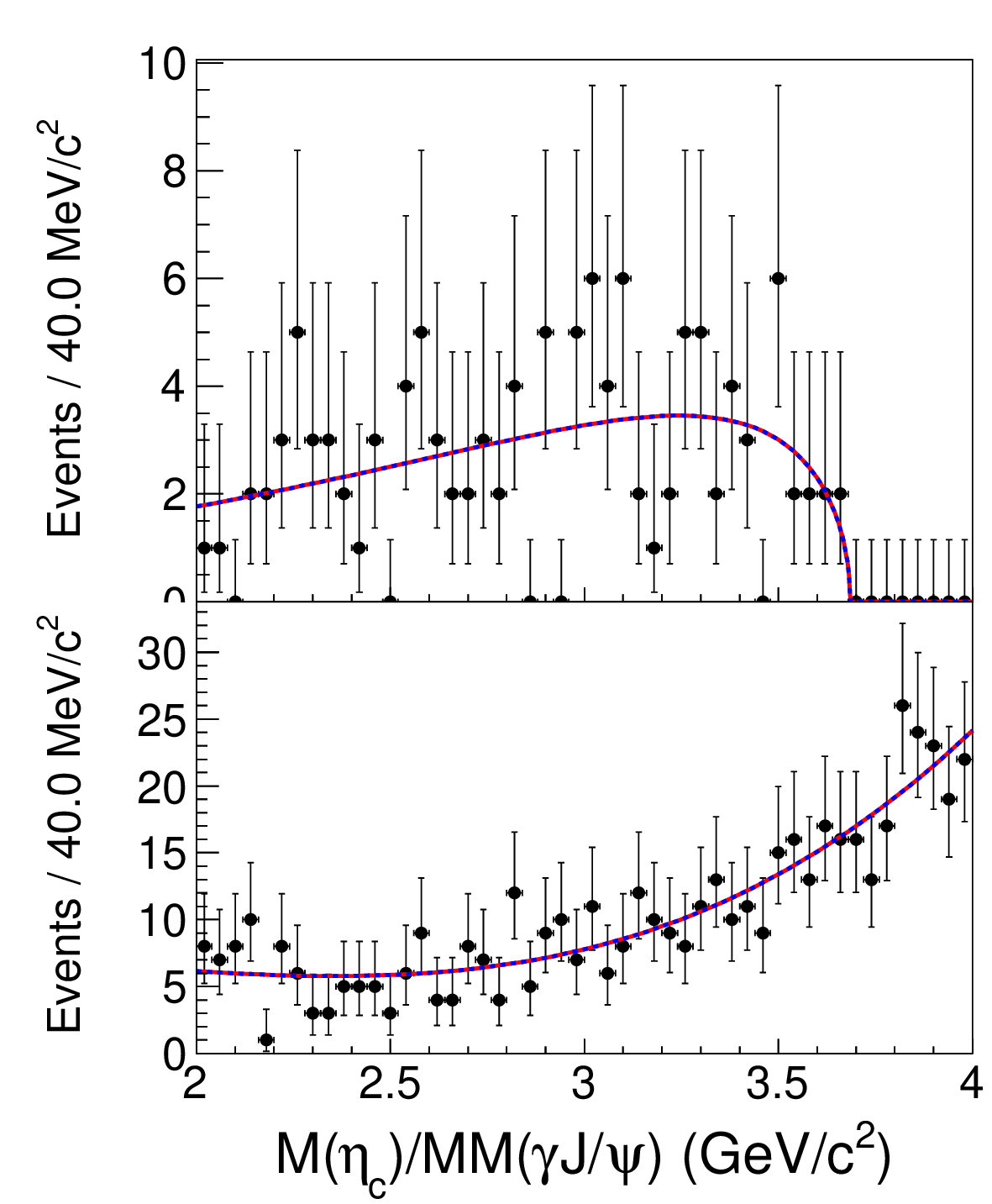}
    \includegraphics[width=0.29\textwidth]{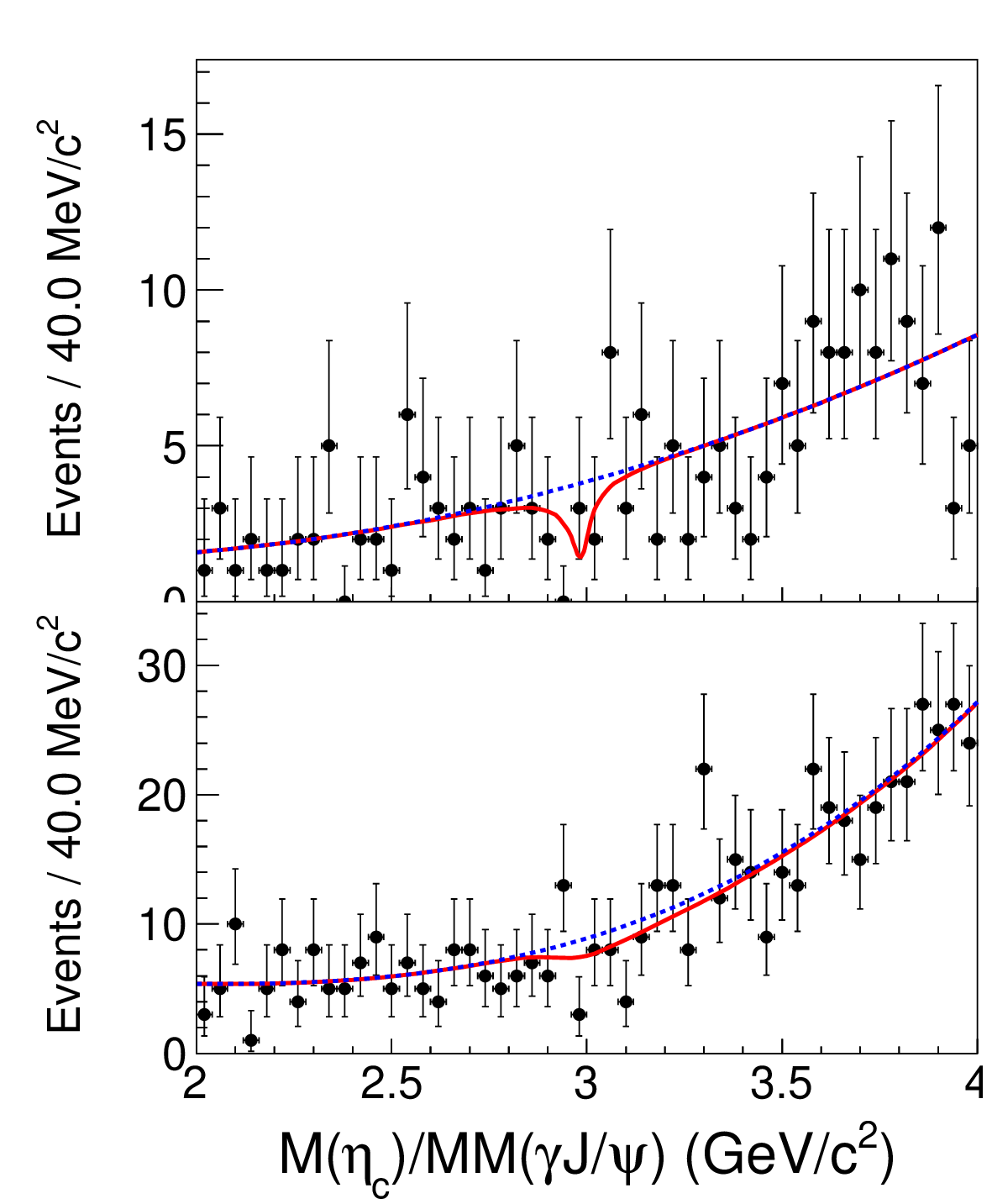}
    \includegraphics[width=0.29\textwidth]{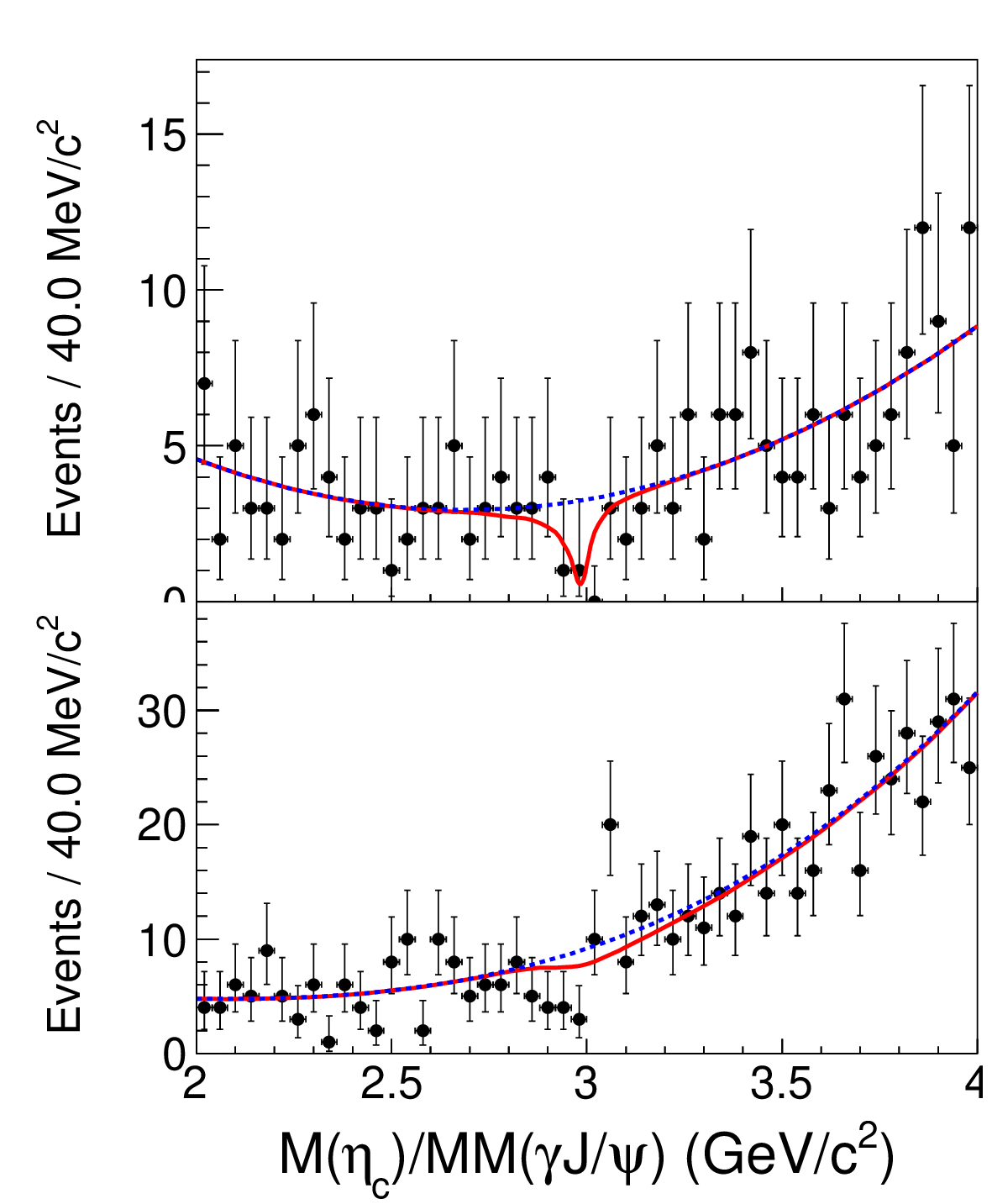} \\
    \includegraphics[width=0.29\textwidth]{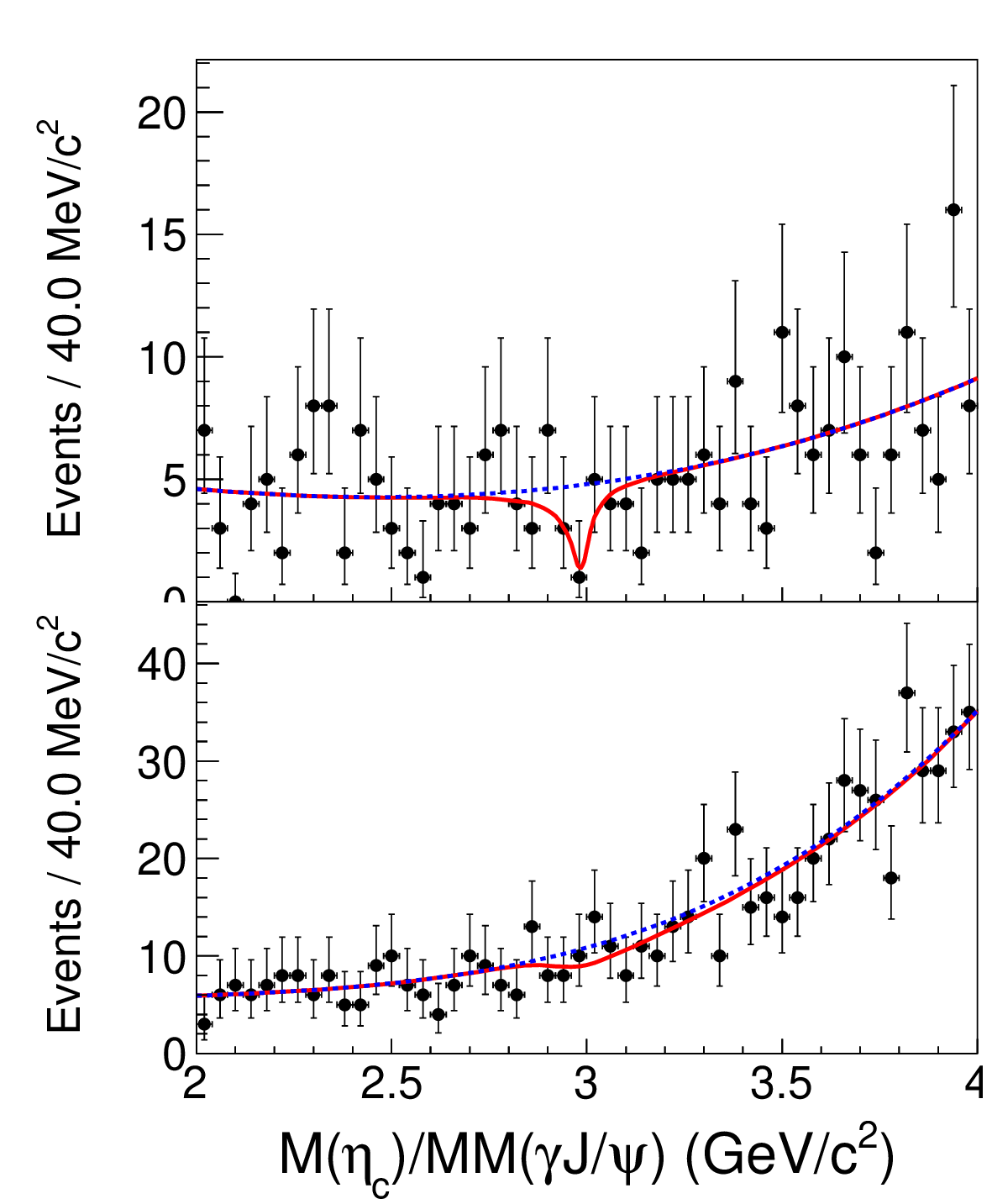}
    \includegraphics[width=0.29\textwidth]{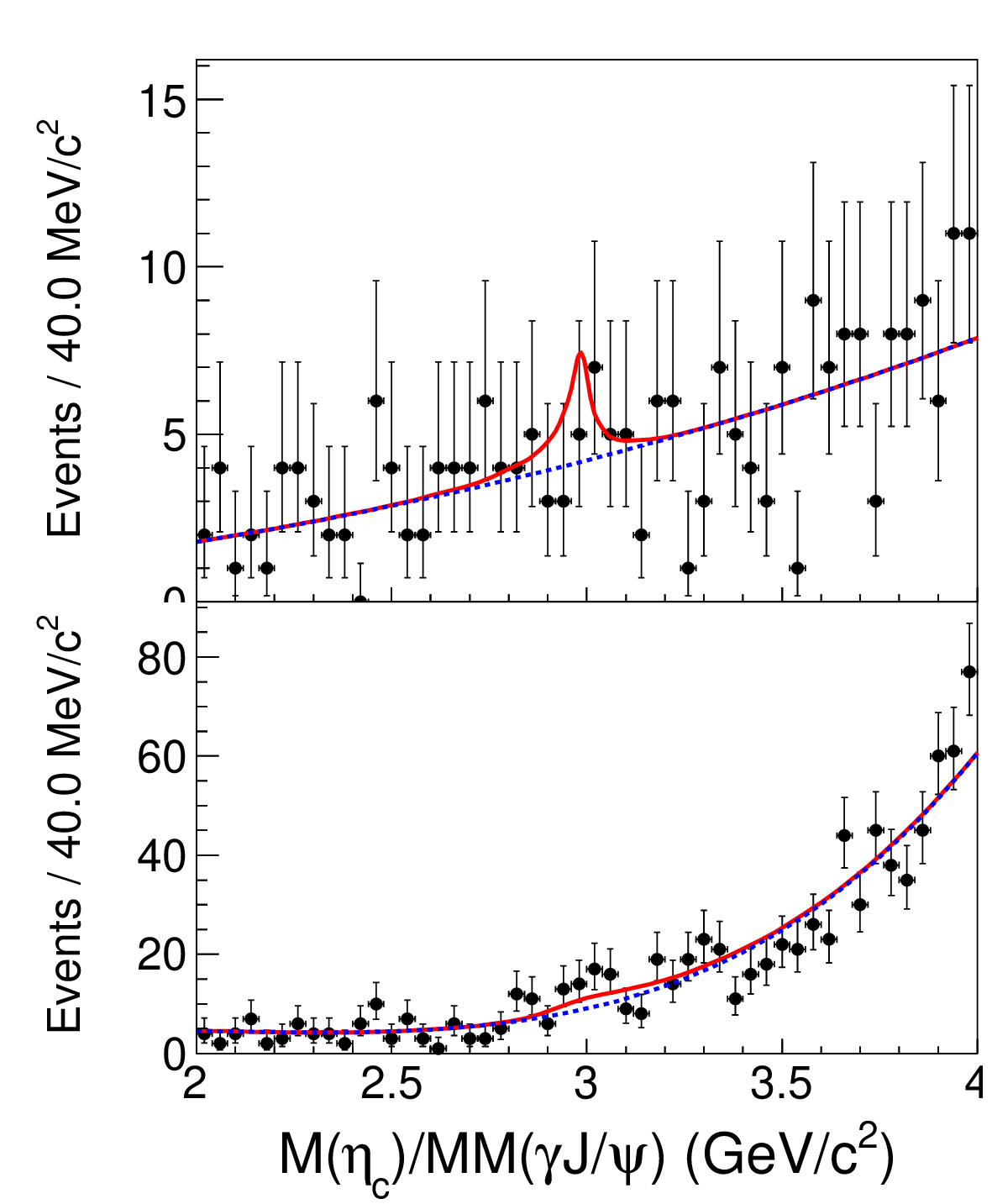}
\end{center}
\caption{Simultaneous fit result to the invariant mass of reconstructed $\etac$ (top) and the $\gisr\jpsi$ recoil mass (bottom). In the top row, from left to right are events with  $M(\etac\jpsi)$ and $M_{\rm recoil}(\gamma)\in[6.4,~6.8],~[6.8,~7.2],~[7.2,~7.6]~\gevcc$, and in the bottom are $\in[7.6,~8.0]$, and [8.0,~8.4]$~\gevcc$. Dots with error bars are from data, the red solid curve is the best fit result, and the blue dashed curve is the background component from the best fit.}\label{fig:SimFit_step400_Etac_2}
\end{figure}

\newpage

\subsubsection{Step size 500 $\mevcc$}


\begin{figure}[htbp]
\begin{center}
    \includegraphics[width=0.23\textwidth]{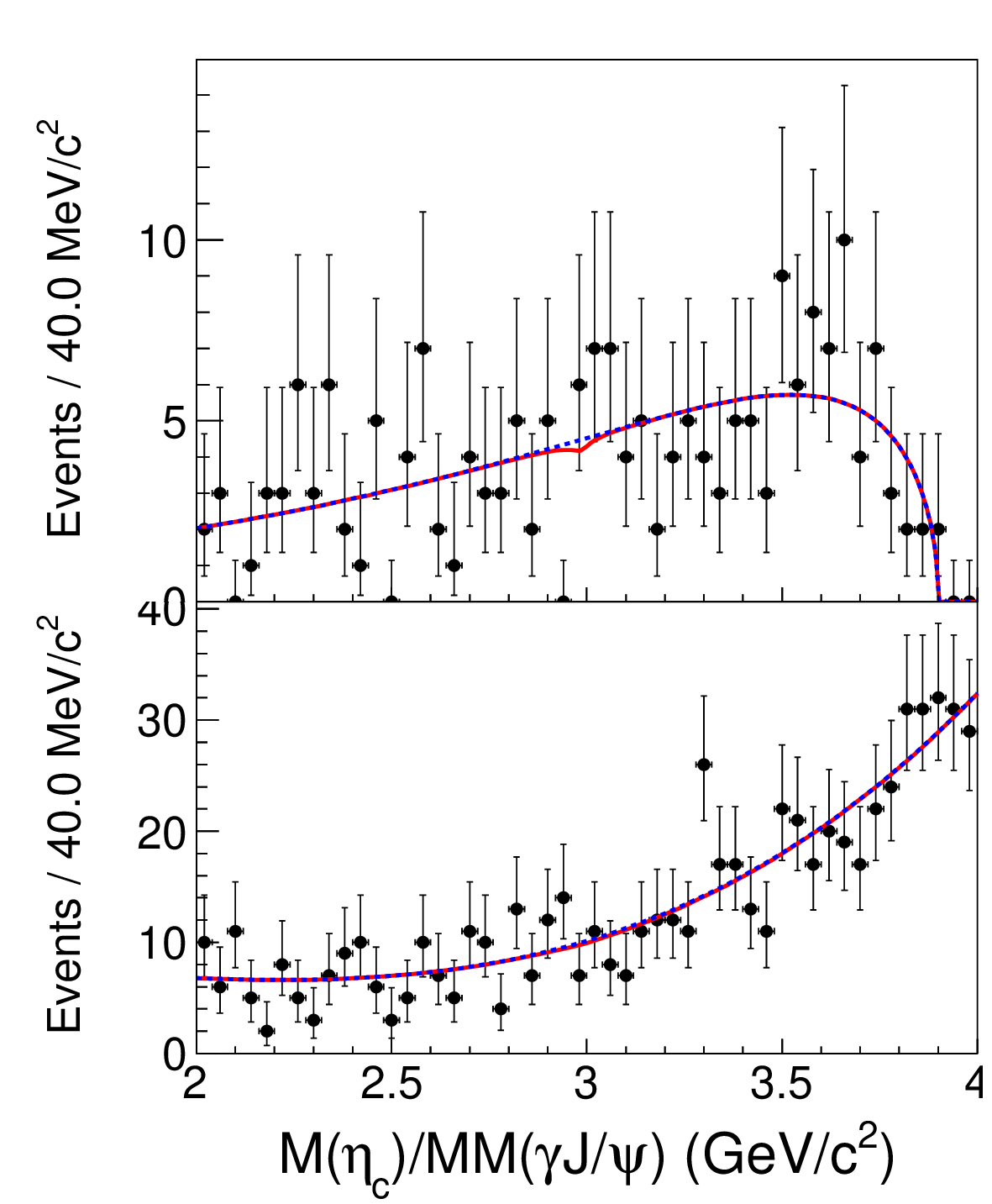}
    \includegraphics[width=0.23\textwidth]{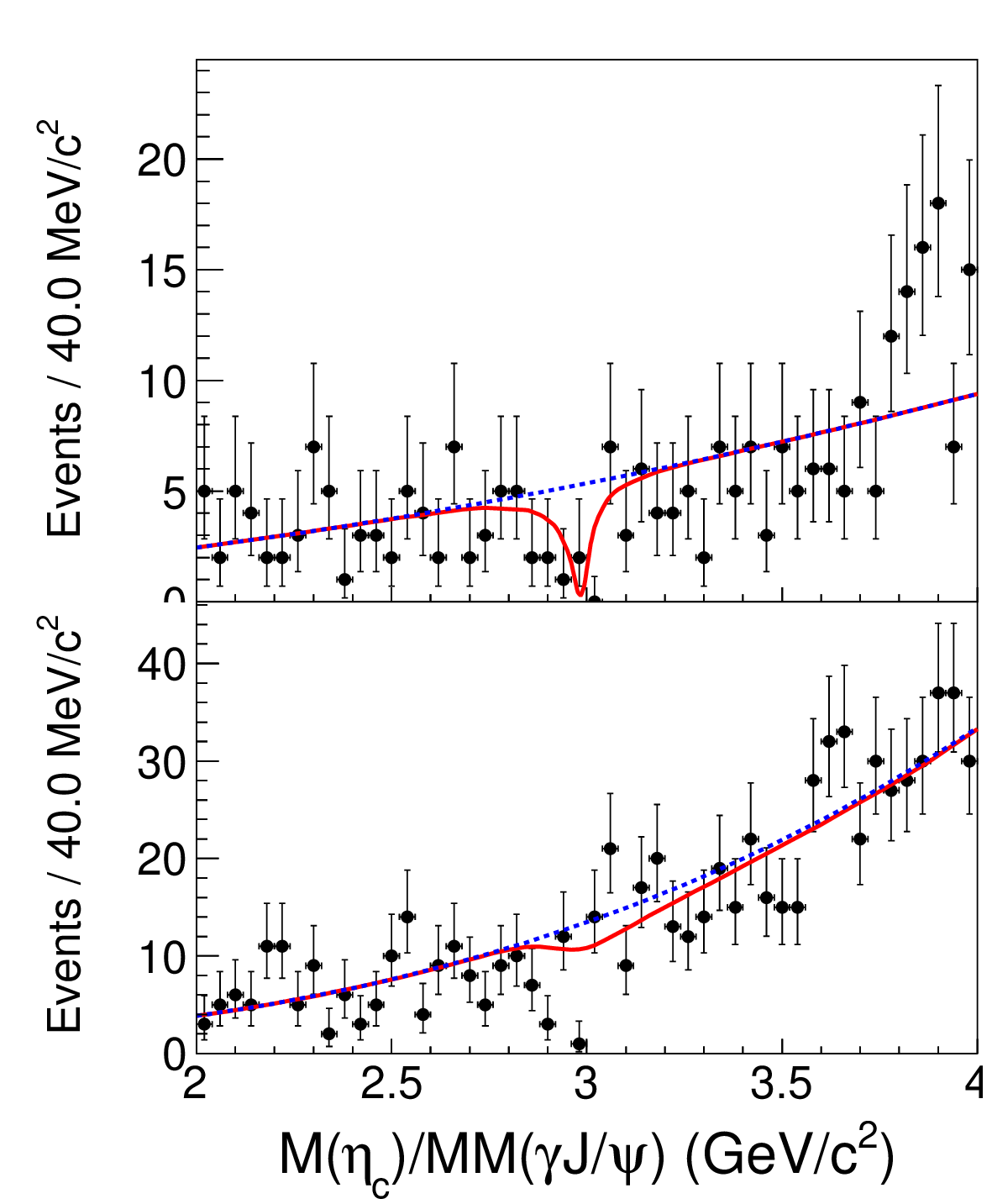} 
    \includegraphics[width=0.23\textwidth]{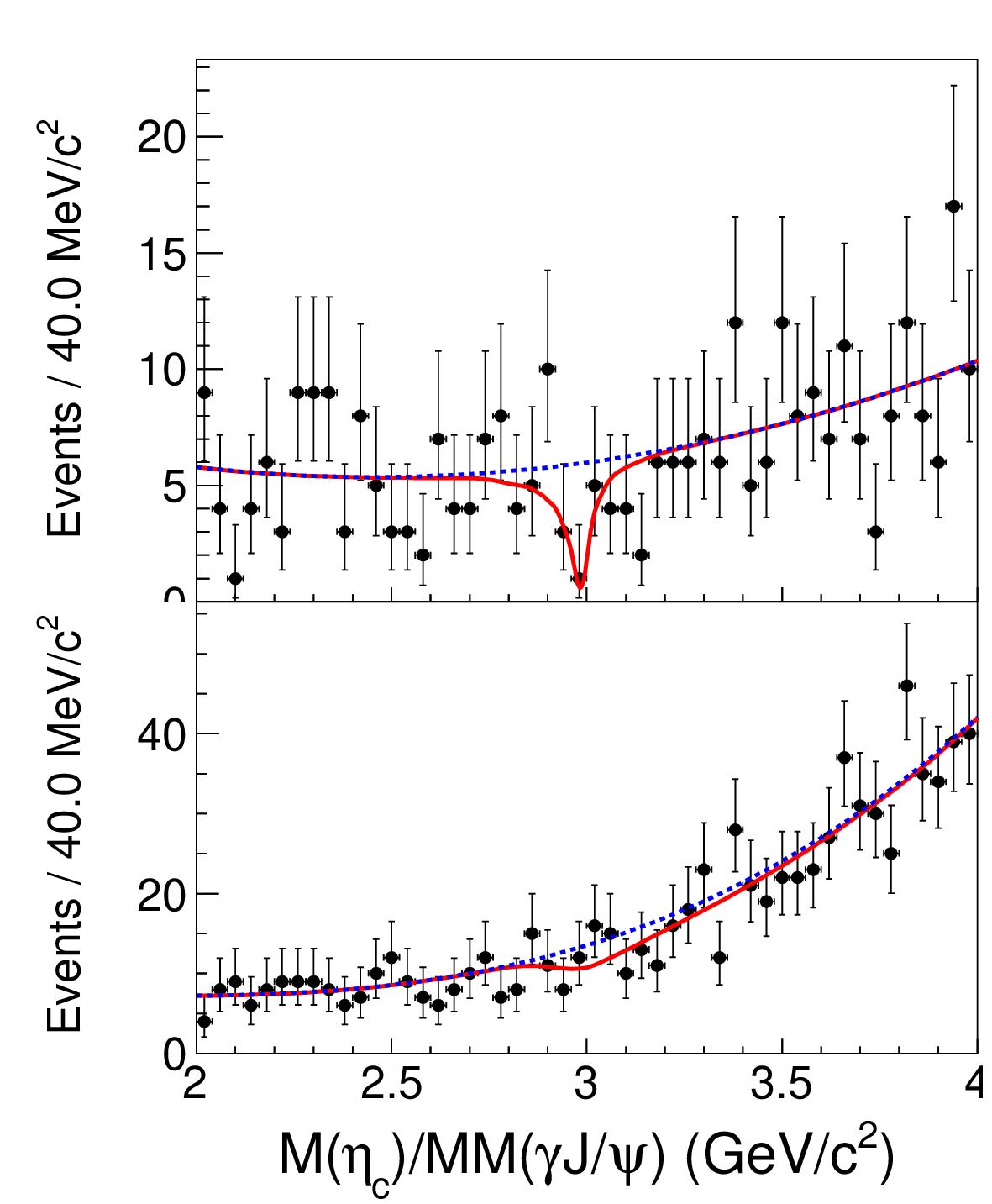}
    \includegraphics[width=0.23\textwidth]{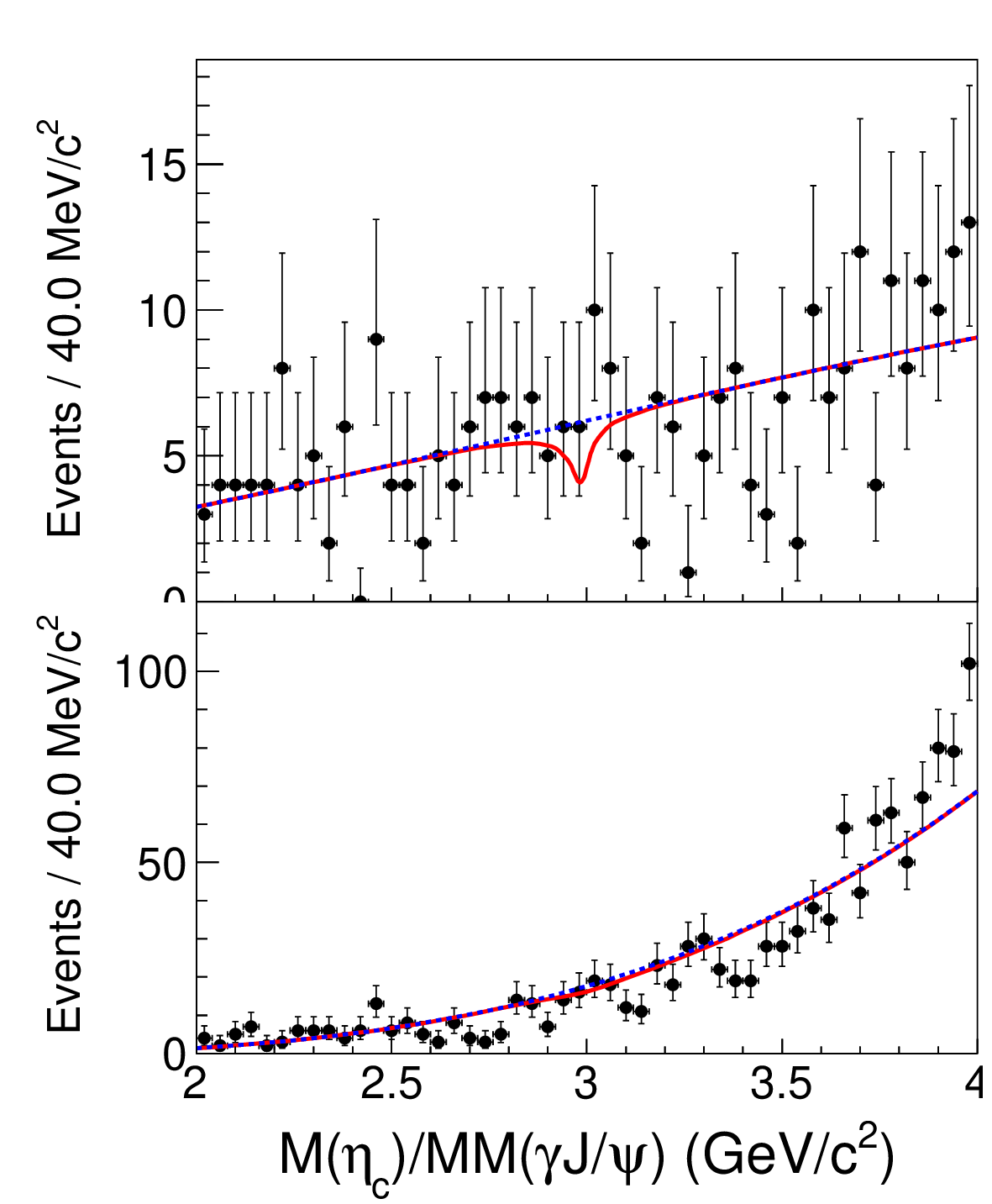}
\end{center}
\caption{Simultaneous fit result to the invariant mass of reconstructed $\etac$ (top) and the $\gamma\jpsi$ recoil mass (right). From left to right are events with  $M(\etac\jpsi)$ and $M_{\rm recoil}(\gamma)\in[6.5,~7.0]$, $[7.0,~7.5]$, $[7.5,~8.0]$, and $[8.0,~8.5]~\gevcc$. Dots with error bars are from data, the red solid curve is the best fit result, and the blue dashed curve is the background component from the best fit.}\label{fig:SimFit_step500_Etac_2}
\end{figure}

\newpage

\subsubsection{Step size 600 $\mevcc$}

\begin{figure}[htbp]
\begin{center}
    \includegraphics[width=0.3\textwidth]{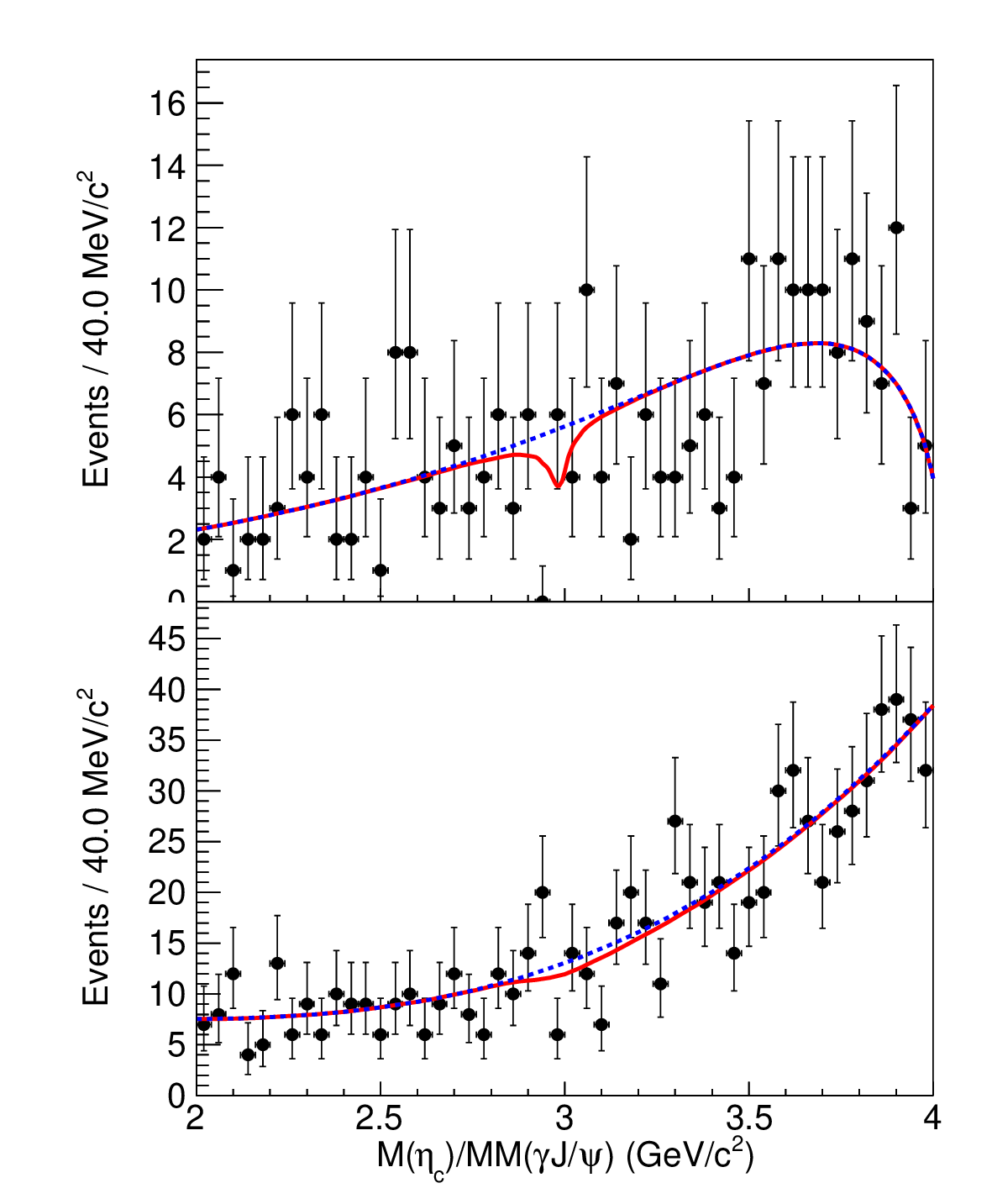}
    \includegraphics[width=0.3\textwidth]{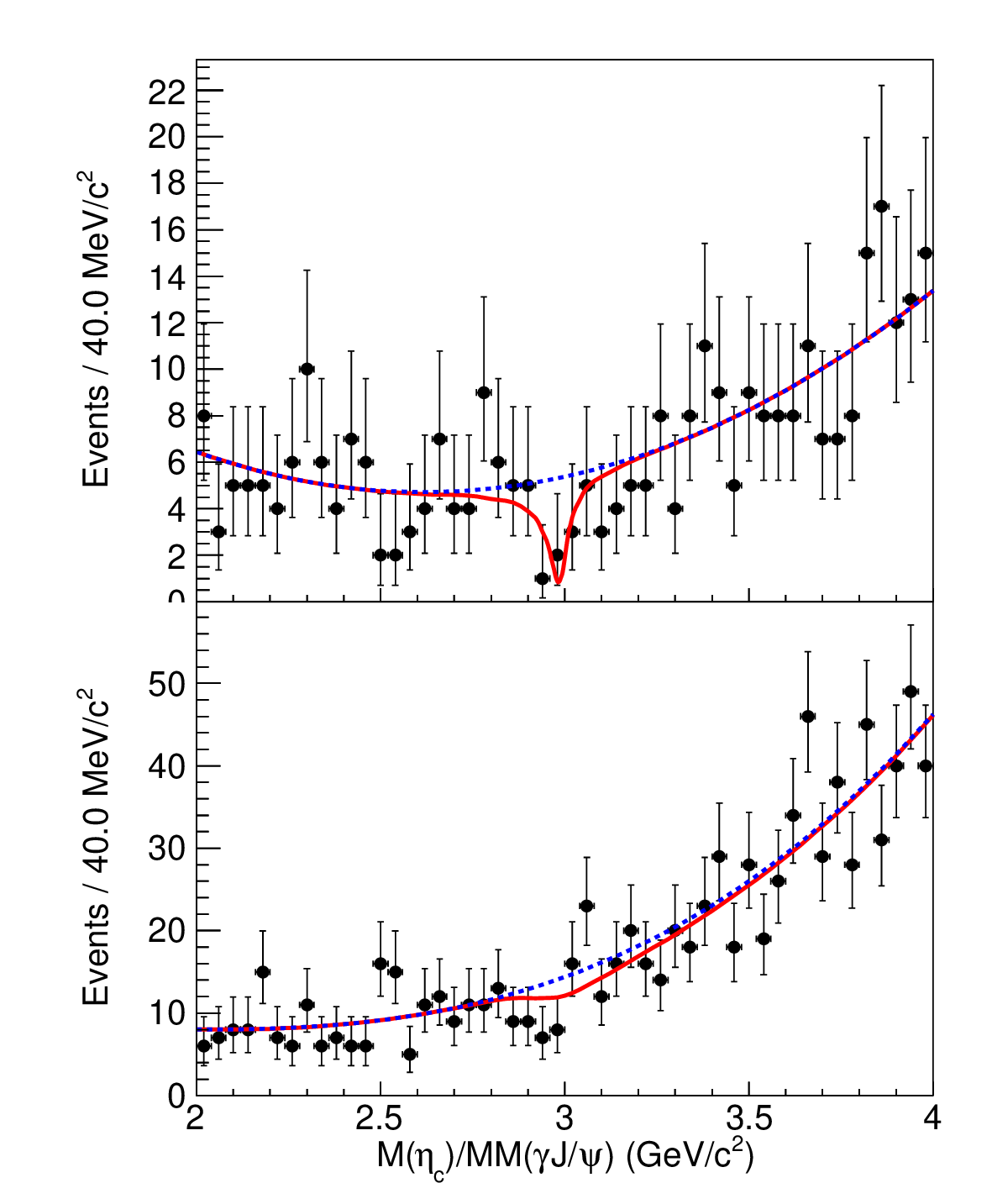}
    \includegraphics[width=0.3\textwidth]{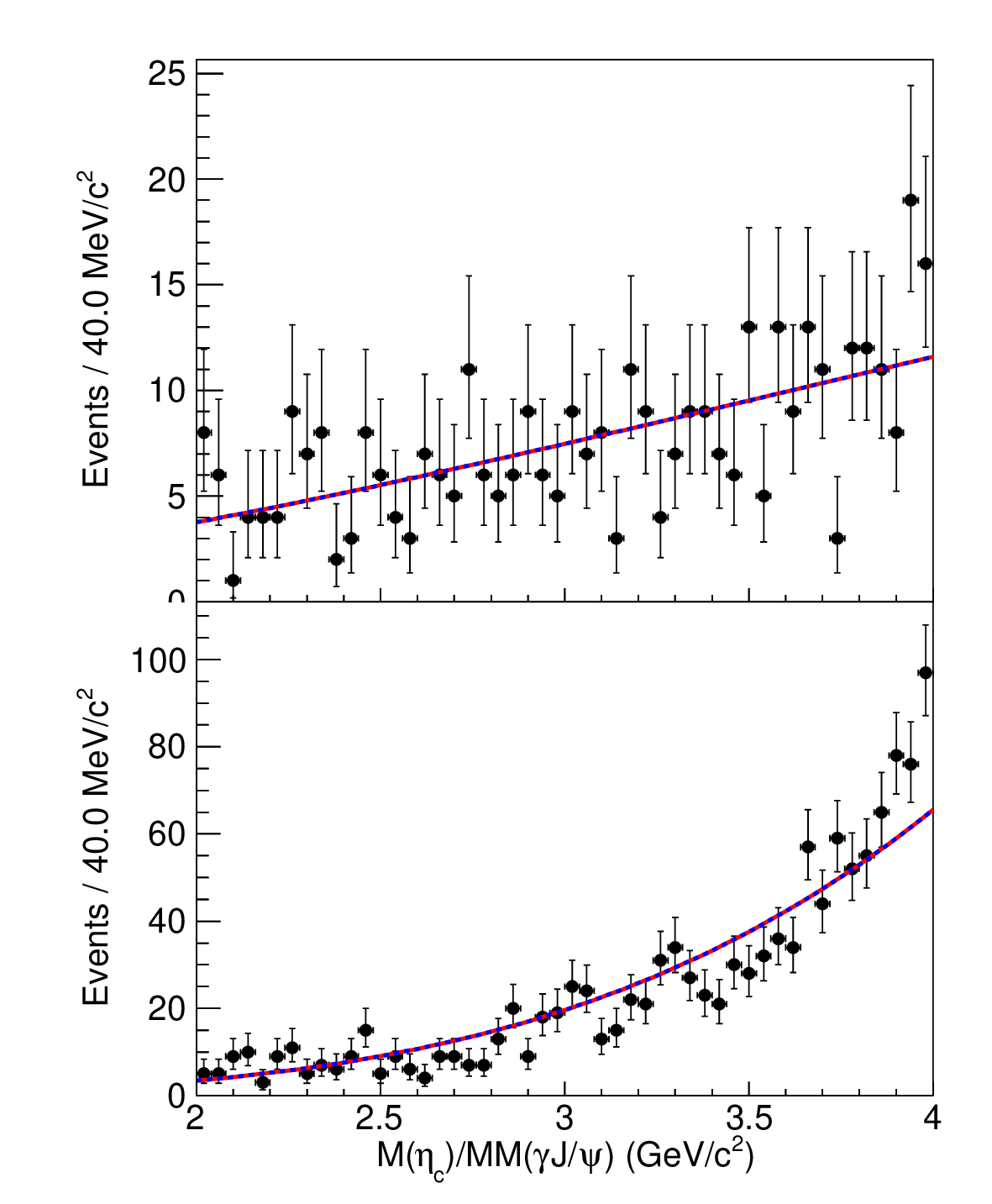}
\end{center}
\caption{Simultaneous fit result to the invariant mass of reconstructed $\etac$ (top) and the $\gamma\jpsi$ recoil mass (right). From left to right are events with  $M(\etac\jpsi)$ and $M_{\rm recoil}(\gamma)\in[6.6,~7.2],~[7.2,~7.8]$, and $[7.8,~8.4]~\gevcc$. Dots with error bars are from data, the red solid curve is the best fit result, and the blue dashed curve is the background component from the best fit.}\label{fig:SimFit_step600_Etac_1}
\end{figure}

\subsection{Systematic uncertainties of fitting procedure}

Here we provide the systematic uncertainties of fitting procedure from different fits.

\begin{table}[htp]
\caption{Uncertainty of the cross sections for different datasets from fit procedure (in unit of \%).}
\begin{center}
\begin{tabular}{c  | c  | c}
\hline
\hline
	dataset		& inclusive     & exclusive	\\
\hline
	 $\Upsilon(1S)$		& 21.5 	&	-		 \\
	 $\Upsilon(2S)$		&  9.8	&	2.2		 \\
	 $\Upsilon(3S)$		&  56.4	&	-		 \\
	 continuum	         & 8.8	&	25.4		  \\
	 $\Upsilon(4S)$ 	& 3.4		&	4.0		 \\
	 $\Upsilon(5S)$		& 13.5	&	16.2		  \\
\hline
\hline
\end{tabular}
\end{center}
\label{tab:sysYnS_fit}
\end{table}%

\begin{table}[htp]
\caption{Uncertainty of the cross sections measurement for different mass regions from fit procedure (in unit of \%).}
\begin{center}
\begin{tabular}{c   c   c c}
\hline
\hline
	regions ($\gevcc$)	& systematic uncertainty  	\\
\hline
	 $[6.0,~6.4]$		& 23.9	 \\
	 $[6.0,~6.5]$		&6.0	 \\
	 $[6.0,~6.6]$		&7.0 \\
	
\hline
\end{tabular}
\end{center}
\label{tab:sysYcc_fit}
\end{table}%


\begin{thebibliography}{99}
\bibitem{reviewXYZ}

N.~Brambilla, S.~Eidelman, C.~Hanhart, A.~Nefediev, C.~P.~Shen, C.~E.~Thomas, A.~Vairo and C.~Z.~Yuan,
Phys. Rept. \textbf{873}, 1-154 (2020).




  \bibitem{Aubert:2005rm}
  B.~Aubert {\it et al.} [BaBar Collaboration],
  Phys.\ Rev.\ Lett.\  {\bf 95}, 142001 (2005).

\bibitem{belle_y4660}
X.~L.~Wang {\it et al.} [Belle Collaboration],
Phys.\ Rev.\ Lett.\ {\bf 99}, 142002 (2007).

\bibitem{babar_y4360}
B.~Aubert {\it et al.} [BaBar Collaboration],
Phys.\ Rev.\ Lett.\ {\bf 98}, 212001 (2007).

\bibitem{cleoY4260}
Q.~He \textit{et al.} [CLEO Collaboration],
Phys. Rev. D \textbf{74}, 091104 (2006).

 \bibitem{Yuan:2007sj}
  C.~Z.~Yuan {\it et al.} [Belle Collaboration],
  Phys.\ Rev.\ Lett.\  {\bf 99}, 182004 (2007).
  
  
\bibitem{Chiu:2005ey}
T.~W.~Chiu \textit{et al.} [TWQCD],
Phys. Rev. D \textbf{73}, 094510 (2006).

  \bibitem{Ablikim:2016qzw}
  M.~Ablikim {\it et al.} [BESIII Collaboration],
  Phys.\ Rev.\ Lett.\  {\bf 118}, 092001 (2017).

  \bibitem{Ablikim:2013wzq}
  M.~Ablikim {\it et al.} [BESIII Collaboration],
  Phys.\ Rev.\ Lett.\  {\bf 118}, 092002 (2017).

\bibitem{czy}
C.~Z.~Yuan,
Chin. Phys. C \textbf{38}, 043001 (2014).

\bibitem{Ablikim:2019apl}
  M.~Ablikim {\it et al.} [BESIII Collaboration],
  Phys.\ Rev.\ D {\bf 99}, 091103 (2019).

\bibitem{Ablikim:2018vxx}
  M.~Ablikim {\it et al.} [BESIII Collaboration],
  Phys.\ Rev.\ Lett.\  {\bf 122}, 102002 (2019).



\bibitem{Y4220etaJpsi}
M.~Ablikim \textit{et al.} [BESIII Collaboration],
Phys. Rev. D \textbf{102}, no.3, 031101 (2020).

\bibitem{Y4220etapJpsi}
M.~Ablikim \textit{et al.} [BESIII Collaboration],
Phys. Rev. D \textbf{101}, no.1, 012008 (2020).


\bibitem{Ablikim:2018epj}
M.~Ablikim \textit{et al.} [BESIII Collaboration],
Phys. Rev. D \textbf{97}, 071101 (2018).



\bibitem{BESIII:2022joj}
M.~Ablikim \textit{et al.} [BESIII Collaboration],
[arXiv:2204.07800 [hep-ex]].

\bibitem{Belle:DsDs1}
S.~Jia \textit{et al.} [Belle Collaboration],
Phys. Rev. D \textbf{100}, no.11, 111103 (2019).

\bibitem{Belle:DsDs2}
S.~Jia \textit{et al.} [Belle Collaboration],
Phys. Rev. D \textbf{101}, no.9, 091101 (2020).

\bibitem{LHCb:phiJpsi1}
R.~Aaij \textit{et al.} [LHCb Collaboration],
Phys. Rev. Lett. \textbf{118}, no.2, 022003 (2017).

\bibitem{LHCb:phiJpsi2}
R.~Aaij \textit{et al.} [LHCb Collaboration],
Phys. Rev. Lett. \textbf{127}, no.8, 082001 (2021).


\bibitem{LHCb_X6900}
R.~Aaij \textit{et al.} [LHCb Collaboration],
Sci. Bull. \textbf{65}, no.23, 1983-1993 (2020).

\bibitem{review_chenhx}
H.~X.~Chen, W.~Chen, X.~Liu, Y.~R.~Liu and S.~L.~Zhu,
Rept. Prog. Phys. \textbf{86} (2023) no.2, 026201
doi:10.1088/1361-6633/aca3b6
[arXiv:2204.02649 [hep-ph]].

\bibitem{cccc_1}
W.~Chen, H.~X.~Chen, X.~Liu, T.~G.~Steele and S.~L.~Zhu,
Phys. Lett. B \textbf{773} (2017), 247-251
doi:10.1016/j.physletb.2017.08.034
[arXiv:1605.01647 [hep-ph]].

\bibitem{cccc_2}
M.~Karliner, S.~Nussinov and J.~L.~Rosner,
Phys. Rev. D \textbf{95} (2017) no.3, 034011
doi:10.1103/PhysRevD.95.034011
[arXiv:1611.00348 [hep-ph]].

\bibitem{cccc_3}
Y.~Bai, S.~Lu and J.~Osborne,
Phys. Lett. B \textbf{798} (2019), 134930
doi:10.1016/j.physletb.2019.134930
[arXiv:1612.00012 [hep-ph]].

\bibitem{cccc_4}
Z.~G.~Wang,
Eur. Phys. J. C \textbf{77} (2017) no.7, 432
doi:10.1140/epjc/s10052-017-4997-0
[arXiv:1701.04285 [hep-ph]].

\bibitem{cccc_5}
V.~R.~Debastiani and F.~S.~Navarra,
Chin. Phys. C \textbf{43} (2019) no.1, 013105
doi:10.1088/1674-1137/43/1/013105
[arXiv:1706.07553 [hep-ph]].

\bibitem{cccc_6}
M.~N.~Anwar, J.~Ferretti, F.~K.~Guo, E.~Santopinto and B.~S.~Zou,
Eur. Phys. J. C \textbf{78} (2018) no.8, 647
doi:10.1140/epjc/s10052-018-6073-9
[arXiv:1710.02540 [hep-ph]].

\bibitem{cccc_7}
A.~Esposito and A.~D.~Polosa,
Eur. Phys. J. C \textbf{78} (2018) no.9, 782
doi:10.1140/epjc/s10052-018-6269-z
[arXiv:1807.06040 [hep-ph]].

\bibitem{cccc_8}
J.~Wu, Y.~R.~Liu, K.~Chen, X.~Liu and S.~L.~Zhu,
Phys. Rev. D \textbf{97} (2018) no.9, 094015
doi:10.1103/PhysRevD.97.094015
[arXiv:1605.01134 [hep-ph]].

\bibitem{cccc_9}
C.~Hughes, E.~Eichten and C.~T.~H.~Davies,
Phys. Rev. D \textbf{97} (2018) no.5, 054505
doi:10.1103/PhysRevD.97.054505
[arXiv:1710.03236 [hep-lat]].

\bibitem{cccc_10}
G.~J.~Wang, L.~Meng and S.~L.~Zhu,
Phys. Rev. D \textbf{100} (2019) no.9, 096013
doi:10.1103/PhysRevD.100.096013
[arXiv:1907.05177 [hep-ph]].
\bibitem{cccc_11}
X.~Chen,
Eur. Phys. J. A \textbf{55} (2019) no.7, 106
doi:10.1140/epja/i2019-12807-2
[arXiv:1902.00008 [hep-ph]].

\bibitem{cccc_12}
M.~S.~Liu, Q.~F.~L\"u, X.~H.~Zhong and Q.~Zhao,
Phys. Rev. D \textbf{100} (2019) no.1, 016006
doi:10.1103/PhysRevD.100.016006
[arXiv:1901.02564 [hep-ph]].

\bibitem{cccc_13}
C.~Deng, H.~Chen and J.~Ping,
Phys. Rev. D \textbf{103} (2021) no.1, 014001
doi:10.1103/PhysRevD.103.014001
[arXiv:2003.05154 [hep-ph]].


\bibitem{belle}A. Abashian {\it et al.} [Belle Collaboration], Nucl. Instrum. Methods Phys. Res., Sect. A 479, 117 (2002); also see detector section in J.~Brodzicka {\it et al.} [Belle Collaboration], PTEP {\bf 2012}, 04D001 (2012).

\bibitem{kekb} S. Kurokawa and E.~Kikutani, Nucl. Instrum. Methods Phys. Res., Sect. A {\bf 499}, 1 (2003), and other papers included in this volume; T. Abe, {\it et al.}, PTEP {\bf 2013}, 03A001 (2013), and following articles up to 03A011.

\bibitem{NNLO-cal}
F.~Feng, Y.~Jia, Z.~Mo, W.~L.~Sang and J.~Y.~Zhang,
[arXiv:1901.08447 [hep-ph]].

\bibitem{Rodrigo:2001kf}
G.~Rodrigo, H.~Czyz, J.~H.~Kuhn and M.~Szopa,
Eur. Phys. J. C \textbf{24} (2002), 71-82
doi:10.1007/s100520200912
[arXiv:hep-ph/0112184 [hep-ph]].

\bibitem{evtgen} D.~J.~Lange, Nucl. Instrum. Meth. A {\bf462}, 152 (2001).

\bibitem{PDG} R.L. Workman {\it et al.} (Particle Data Group), Prog. Theor. Exp. Phys. 2022, 083C01 (2022).


\bibitem{geant3} R. Brun et al., CERN Report No. DD/EE/84-1 (1987).
\bibitem{topoana} X.~Zhou, S.~Du, G.~Li and C.~Shen, Comput. Phys. Commun. \textbf{258}, 107540 (2021).

\bibitem{PID}
E.~Nakano, Nucl.\ Instrum.\ Methods Phys.\ Res.\ Sect. A {\bf 494}, 402 (2002).

\bibitem{muID} A.~Abashian \textit{et al.}, Nucl.\ Instrum.\ Methods Phys.\ Res.\ Sect.\ A {\bf 491}, 69 (2002).

\bibitem{eID} K.~Hanagaki, H.~Kakuno, H.~Ikeda, T.~Iijima, and T.~Tsukamoto,
Nucl.\ Instrum.\ Methods Phys.\ Res.\ Sect.\ A {\bf 485}, 490 (2002).


\bibitem{NNKs}H. Nakano, Ph.D Thesis, Tohoku University (2014)
Chapter 4, http://hdl.handle.net/10097/58814.

\bibitem{rookeyspdf} K.~S.~Cranmer, Comput. Phys. Commun. \textbf{136}, 198-207 (2001).


\bibitem{trolke} The upper limit is calculated by using a frequentist method with unbounded profile likelihood treatment of systematic uncertainties, which is implemented by a C++ class {\sc trolke} in the {\sc root} framework~\cite{{upcal}}. The number of the observed events is assumed to follow a Poisson distribution, the number of background events and the efficiency are assumed to follow Gaussian distributions. 

\bibitem{upcal}W.~A.~Rolke, A.~M.~Lopez and J.~Conrad,
Nucl. Instrum. Meth. A \textbf{551}, 493-503 (2005). 

\bibitem{BelleCC}
K.~Abe \textit{et al.} [Belle Collaboration],
Phys. Rev. Lett. \textbf{98}, 082001 (2007).

\bibitem{BelleCC2}
S.~D.~Yang \textit{et al.} [Belle Collaboration],
Phys. Rev. D \textbf{90}, no.11, 112008 (2014).


\bibitem{12Smm}
M.~Kobel \textit{et al.} [Crystal Ball Collaboration],
Z. Phys. C \textbf{53}, 193-206 (1992).



\bibitem{fwmomentum} The Fox-Wolfram moments were introduced by G. C. Fox
and S. Wolfram in Phys. Rev. Lett. {\bf 41}, 1581 (1978). The
modified Fox-Wolfram moments used in this paper are
described in S. H. Lee et al. [Belle Collaboration], Phys. Rev. Lett. {\bf 91}, 261801 (2003).


\bibitem{ARGUS}
  H.~Albrecht {\it et al.} [ARGUS Collaboration],
  Phys.\ Lett.\ B {\bf 241}, 278 (1990).

\bibitem{effLumi}
M.~Benayoun, S.~I.~Eidelman, V.~N.~Ivanchenko and Z.~K.~Silagadze,
Mod. Phys. Lett. A \textbf{14}, 2605-2614 (1999)
doi:10.1142/S021773239900273X
[arXiv:hep-ph/9910523 [hep-ph]].


\bibitem{kserr} N.~Dash, S.~Bahinipati, V.~Bhardwaj, K.~Trabelsi \textit{et al.}
[Belle Collaboration]
Phys. Rev. Lett. \textbf{119}, 171801 (2017).

\bibitem{pi0err} S.~Ryu \textit{et al.} [Belle Collaboration],
Phys. Rev. D \textbf{89}, 072009 (2014).


\bibitem{BelleLumi} G. Rodrigo {\it et al.}, Eur. Phys. J. C 24, 71 (2002). For a review on the generator, see: S. Actis {\it et al.}, Eur. Phys. J. C 66, 585 (2010).

\bibitem{ISRtheory}
  E.~A.~Kuraev and V.~S.~Fadin,
  Sov.\ J.\ Nucl.\ Phys.\  {\bf 41}, 466 (1985).



\newpage

\end{thebibliography}
\end{document}